\documentclass[aps,prd, 
amsmath,amssymb,showpacs,nofootinbib]{revtex4}
\usepackage{graphicx}

\usepackage[usenames,dvipsnames]{color}
\usepackage{hyperref}
\usepackage{color}

\newbox\bwk\edef\tempd#1pt{#1\string p\string t}
\tempd\def\nbextr#1pt{#1}

\def\npts#1{\expandafter\nbextr\the#1\space}
\def\ttwplink#1#2{
#2
\setbox\bwk=\hbox{#2}\special{ps:( linkto #1)\space\npts{\wd\bwk} \npts{\dp\bwk} -\npts{\ht\bwk} true\space Cpos}}
\newcommand{\cm}[1]{}
\newcommand{\eq}[1]{\begin{equation}#1\end{equation}}
\newcommand{\secs}[1]{\section{#1\label{sec-#1}}}
\newcommand{\ssecs}[1]{\subsection{#1\label{ssec-#1}}}

\newcommand{\eqa}[1]{\begin{eqnarray}#1\end{eqnarray}}
\newcommand{\fig}[4]{\begin{figure}[htbp]\centering\includegraphics[width=#3\textwidth]{#1}\caption{#2}\label{fig-#1}\end{figure}}
\newcommand{\tab}[2]{\begin{table}\centering\caption{#1}#2\label{tab-#1}\end{table}}
\newcommand{\subs}[1]{_\mathrm{#1}}

\newcommand{\dd}[1]{\mathrm{d}#1}
\newcommand{\mpl}{M\subs{p}}
\newcommand{\refeq}[1]{Eq.\ \ref{eq-#1}}
\newcommand{\reftab}[1]{Table\ \ref{tab-#1}}
\newcommand{\refcn}[1]{Constraint\ \ref{cnstr-#1}}
\newcommand{\refig}[1]{Fig.\ \ref{fig-#1}}
\newcommand{\refsec}[1]{Section \ref{sec-#1}}
\newcommand{\refssec}[1]{Section \ref{ssec-#1}}

\newcommand{\ta}[1]{\langle #1\rangle}

\newcommand{\ms}{M\subs{s}}

\newcommand{\add}[1]{{\Large #1}}
\newcommand{\del}[1]{{\scriptsize(#1)}}

\cm{
\newcommand{\add}[1]{#1}
\newcommand{\del}[1]{}

}

\newtheorem{constraint}{{\footnotesize CONSTRAINT}}
\newcommand{\cnstr}[2]{\renewcommand{\l}{\vspace{0.1in}\\\displaystyle }\begin{constraint}[#1\label{cnstr-#1}]\eq{\begin{array}{r@{,\hspace{0.5in}\mathrm{for}\hspace{0.1in}\displaystyle }l}\displaystyle #2\end{array}}\end{constraint}}

\def\VEV{{\it{vev}}}
\def\vev{{\it{vev}}}
\newcommand{\void}[1]{}
\begin{document}
\preprint{arXiv:1101.0202 [astro-ph:CO]\cr
NITS-PHY-2011001}

\title{Bound To Bounce: A Coupled Scalar-Tachyon Model For A Smooth Bouncing/Cyclic Universe}

\author{Changhong~Li,}
\email[]{chellifegood@gmail.com}
\author{Lingfei ~Wang,}
\email[]{W0F000@gmail.com}
\author{Yeuk-Kwan~E.~Cheung}
\email[Correspondance:]{cheung@nju.edu.cn \& cheung.edna@gmail.com}
\affiliation{Department of Physics, Nanjing University\\
22 Hankou Road, Nanjing, China 210093}

\begin{abstract}
We introduce a string-inspired model for a bouncing/cyclic universe, utilizing the scalar-tachyon coupling as well as contribution from curvature in a closed universe. 
The universe undergoes the locked inflation, tachyon matter dominated rolling expansion, turnaround and contraction, as well as the subsequent deflation and ``bounce'' in each cycle of the cosmological evolution.  
We perform extensive analytic and numerical studies of the above evolution process. 

The minimum  size of the universe is nonzero for generic initial values.  The smooth bounce are made possible  because of the negative contribution to effective energy density by the curvature term.  
No ghosts are ever generated at any point in the entire evolution of the universe,   
with the Null, Weak, and Dominant Energy Conditions preserved even at the bounce points, contrary to many bounce models previously proposed.  
And the Strong Energy Condition is satisfied in periods with tachyon matter domination. 

\end{abstract}

\maketitle

\tableofcontents

\secs{Introduction}

The idea of modelling the exponential expansion of the universe using simple scalar fields has led to remarkable progress in the study of cosmology~\cite{Guth:1980zm, Starobinsky:1980te, Sato:1980yn}, crystalizing the earlier attempts~\cite{Zeldovich:1972zz, Sunyaev:1970eu, Brout:1977ix, Starobinsky:1979ty}--notably by Zeldovish and by Starobinsky--and paving the way to later developments~\cite{Kazanas:1980tx, Sato:1981ds, Starobinsky:1982ee, Linde:1981mu,Albrecht:1982wi, Linde:1986fc, Linde:1993cn}. In particular, the mechanism for generating matter from quantum fluctuations of the scalar field~\cite{Mukhanov:1990me} successfully endows the inflation scenario with a casual mechanism for generating  matter density fluctuations that agrees spectacularly well with the current array of observations~\cite{Smoot:1992td, Bennett:1996ce, Spergel:2006hy,Komatsu:2008hk}.   This success has been crowned the ``inflation paradigm'' \cite{Liddle:2000cg, Mukhanov:2005sc}.    
Despite the success story, improved theoretical understanding, on the other hand, points to many shortcomings of the paradigm.  The most obvious is the lack of a quantum theory of gravity in modelling physics of  the early universe.  This is in turn reflected by the fact that the effective theory using scalar field necessarily breaks down as one approaches the Planck scale, $\mpl$,  and cosmic singularities are inevitable~\cite{Borde:1993xh}. And trans-Plankian effects will eventually show up in the late time physics~\cite{Martin:2000xs, Easther:2001fi}. See, however,~\cite{Starobinsky:2001kn, Starobinsky:2002rp,Burgess:2002ub, Schalm:2004xg} for a different school of thought.  

In  an  attempt to address the Big Bang singularity,  various alternative models are proposed.  These models can be loosely divided into  four categories. The first consists of models that  implement T-duality to circumvent the Big-Bang singularity,  such as  
Brandenberger-Vafa scenario~\cite{Brandenberger:1988aj, Patil:2004zp, Brandenberger:2005bd, Kaloper:2006xw}, string/brane gas bouncing universe models~\cite{Greene:2008hf} and pre-big-bang models~\cite{Gasperini:2002bn}. Models with modifications to Einstein's theory of General Relativity (GR) to make gravity asymptotically free at very high energy, the like of a bouncing universe using  $f(R)$ gravity~\cite{Biswas:2005qr} belong to the second category. 
The third is comprised of D-brane collisions in extra dimension  as the interpretation of the Big-bang, building on the idea of~\cite{Dvali:1998pa}.  
Examples are Ekpyrotic universe~\cite{Khoury:2001wf} and cyclic Ekpyrotic scenarios~\cite{Steinhardt:2001st}. See~\cite{Shtanov:2002mb} as an example of other attempts to solve singularity in brane-world.  
The fourth is standard GR based bouncing models in which the dynamic behavior of each component's equation of state is applied to the  bounces,  such as the bouncing universe with quintom matter~\cite{Cai:2007qw, Cai:2008qb, Brandenberger:2009ic, Cai:2007zv}.  Models in this category received a lot of attention as one can actually write down a Lagrangian and solve for the dynamic equations, even though  many of these attempts encountered ghosts  at  bounce points.  

In this paper, we introduce a string-inspired coupled scalar-tachyon fields model in the closed FLRW background.  The newly introduced scalar-tachyon coupling term  make the coupled tachyon condensation come out naturally after the locked inflation driven by the tension of D-anti-D brane pairs. Further investigation reveals that such scalar-tachyon coupling term plays a crucial role in cosmic evolution of universe, and distinguishes the coupled scalar-tachyon model from other traditional single tachyon field cosmology~\cite{Sen:2002in,Sen:2002nu,Gibbons:2002md} and/or D-brane inflation cosmology~\cite{HenryTye:2006uv,Dvali:1998pa,Dvali:2001fw,Burgess:2001fx}.
Accordingly, the cosmological evolution of this model consists of six distinctive phases: locked inflation, coupled tachyon condensation, tachyon matter dominated era(expansion, turnaround and contraction), reversal tachyon condensation, deflation and smooth bounce. The bounce process is sufficiently soft , $\ddot{a}_\ast\le a_\ast^{-1}$. And the Null Energy Condition, Dominant Energy Condition and Weak Energy Condition are not ever violated, which is consistent to a model-independent analysis in~\cite{MolinaParis:1998tx}. No ghosts are ever generated because of the kinetic terms of each field in this model take correct sign, $+$.

This paper is organized as following. In \refsec{MB}~, we establish the coupled scalar-tachyon fields model by considering the unification of D-brane inflation and tachyon condensation process, then give a overview of the comic evolution of this model in closed FLRW background.  Each stage of universe evolution is studied in detail in \refsec{cem}~, including analytical analysis and numerical simulation. In \refsec{concl}~, we conclude this paper.

\section{The Coupled Scalar-Tachyon Field Bouncing Universe Model}
\label{sec-MB}
Tachyon was taken as an unphysical object historically. However, the modern quantum field theory revived the concept of tachyon by offering a correct interpretation that, the existence of a tachyon signals the emergence of an  instability of the quantum system, and such instabilities used to be associated with the phase transitions of symmetries breaking
\footnote{
A tachyon was defined,  in the past,  as a particle that travelled  faster than light, which implied the mass-square of tachyon  was negative. Such ``faster-than-light'' feature rendered tachyon utterly unphysical until quantum field theory provides  a correct understanding of tachyon concept. In quantum field theory, the mass-square of a particle-like state  is introduced as $m^2=\frac{d^2 V(\chi)}{d\chi^2}|_{\chi=0}$, where $\chi$ is a scalar field, $V(\chi)$ is its potential and has an extremum at the origin. It is clear that for $V^{\prime\prime}(0)<0$ the particle of such state has a negative mass-square, which is nothing but  a tachyon mentioned above. In this context, the existence of tachyon has a reasonable physical interpretation: 1) For $V^{\prime\prime}(0)>0$, it  describes a particle with positive mass-square. And the extremum of the potential $V(\chi)$ at the origin is a minimum. With small displacement of $\chi$, $\chi$  oscillates around the origin. The system is stable. 2) For $V^{\prime\prime}(0)<0$, it describes a tachyon, and the extremum of the potential $V(\chi)$ at the origin is a maximum. Therefore, with small perturbation, the $\chi$ will roll down from the origin and grow rapidly in time. The existence of a tachyon, therefore, signals the emergence of an  instability of the quantum system. Of course, the tachyon should be always understood as a sort of particle-like state with negative ``effective'' mass rather than some elementary particles. A example of tachyon field in Standard Model~\cite{Peskin:1995ev} is that, at the moment electroweak symmetry $SU(2)\times U(1)$ breaking, the Higgs field is becoming tachyonic and rolling from origin of its field space, which is the minimum of Higgs potential before the electroweak symmetry breaking, to the true vacuum after the electroweak symmetry breaking. This observation unveils an encouraging fact that the emergence of tachyon does not necessarily imply a flaw in this theory.  In contrary, the study of the tachyonic behaviors of the field can offer us a better understanding of the dynamics of phase transition related to symmetry breaking.}.
In superstring theories, the open string tachyon widely appears in various non-supersymmetric structures, such as non-BPS Dp-branes and  the BPS Dp--anti-Dp brane pair~\cite{Polchinski:1996na, Banks:1995ch, Sen:1998ii, Bergman:1999kq}. And the open string tachyon signals the instability of these D-branes and/or anti-D-branes structures.  It turns out these non-supersymmetric structures play a crucial role in understanding the descent relations among different dimensional D-branes and the annihilation and production of D-branes~\cite{Ghoshal:2000dd,Sen:1999mh}.  Utilizing the purely string theory techniques, Ashoke Sen and other researchers~\cite{Sen:1998sm,Sen:1999nx, Sen:2002in, Sen:2002nu} show that the dynamics of open string tachyon can depict the decay process of these unstable non-BPS D-branes and BPS D--anti-D-branes pair. Similar to electroweak symmetry breaking in which the Higgs field becomes tachyonic, open string tachyons appear when supersymmetry is being broken.  The open string tachyon condensation,  henceforth, describes the phase transition after the  breaking of supersymmetry of this structures, see a nice review for this issue~\cite{Sen:2004nf}. Building upon this modern knowledge of open string tachyons in various superstring theories,  cosmological implications of tachyon condensation  have  subsequently  been explored in a large number of works, for example see~\cite{Gibbons:2002md, Fairbairn:2002yp, Feinstein:2002aj, Mazumdar:2001mm, Bagla:2002yn, Shiu:2002qe}.

In this section, we establish a coupled scalar-tachyon fields model based on unstable D-brane theory, and give an overview of this model in closed FLRW background.  The detailed analysis--analytic as well as numerical--for each stage of the cosmic evolution of universe according to this model are presented in coming sections.

\subsection{Model Building}
\label{sec-Building}
In this paper, we specialize in the type IIB string theory with a coincident BPS D3--anti-D3 brane pair. In effective theory, such D3--anti-D3 brane pair is described by the single open string tachyon field action~\cite{Gibbons:2002md, Sen:2003mv, Garousi:2000tr}:
\begin{equation}
S=\int\dd^4xV(T)\sqrt{1+\ms^{-4}\partial_\mu T\partial^\mu T},\hspace{0.45in}V(T)=\frac{V_0}{\cosh(\frac{T}{\sqrt{2}\ms})}~, \label{eq-sta1}
\end{equation}
where $T$, $\ms$, $V_0$ are the tachyon field, the string mass, the tension of D3--anti-D3-branes pair respectively. The tachyon potential $V(T)$ has a maximum at $T=0$ and two minima at $T \rightarrow\pm\infty$. We take the metric as $(-,+,+,+)$. And we also assume the tachyon field is spatially homogenous. 

In the single tachyon field model, initially tachyon field is localized on the top of its potential, $\dot{T}=0$ and $T=0$ while the pair of static D3-anti-D3 branes lay over each other. Such initial state forms an open string vacuum in string theory~\cite{Sen:2004nf}. However, the open string vacuum is unstable. With a small perturbation, tachyon field is rolling down from the top of its potential, and the $T$ and $\dot{T}$ grow swiftly toward $M_s^2t$ and $M_s^2$.  This tachyon rolling process signals the annihilation of the static D3-anti-D3 branes. After D3-anti-D3 brane pair annihilating, only matter-like relics are left. This whole process is called as ``tachyon condensation'' in the single tachyon field theory.

This single tachyon condensation can be viewed as a phase transition from dark energy-like component to matter-like component. According to \refeq{sta1}~, the Equation of State of tachyon field is that, $\omega_T=-(1-\ms^{-4}\dot{T}^2)$~. Before tachyon condensation taking place, $\dot{T}=0$, tachyon field behaves as dark energy, $\omega_T=-1$. It indicates the tachyon field behaves like dark energy. Such ``tachyon dark energy'' is just the tension of the static D-anti-D-branes pair. After tachyon condensation, $\dot{T}\rightarrow M_s^2$, D-anti-D-branes pair annihilates and tachyon field condensates to be tachyon matter, $\omega_T\rightarrow0$.

On the other hand,  in the D-brane inflation scenario~\cite{Dvali:1998pa,Dvali:2001fw}, the D3 and anti-D3 branes are separated by a long distance $y$, which is much larger than $M_s^{-1}$. There is an attractive force, between the D3-brane and anti-D3-brane due to the exchange of closed string between these two branes. This attractive potential takes following form~\cite{HenryTye:2006uv}:
\begin{equation}
V_\phi=\frac{1}{2}m^2\phi^2+V_0-\frac{V_0^2}{4\pi^2 v \phi^4}~, \quad \phi\equiv\sqrt{V_0}y~.\label{eq-Vphi}
\end{equation}
Due to this attractive potential, universe inflates when these two branes move forward each other. In general, D-brane inflation ends with the total annihilation of the  two branes, resulting in the condensation of the tachyon. Models of this kind are derived from string cosmology, and have been widely studied (See~\cite{Quevedo:2002xw} for a review.) since the pioneer work of Dvali, Tye and other researchers~\cite{Dvali:1998pa, Dvali:2001fw, HenryTye:2006uv, Burgess:2001fx}.  In these models, the annihilation process of these two branes serves as the exit of D-brane inflation.

We are however motivated to consider the combined effect of tachyon condensation and D-brane inflation. By introducing a scalar-tachyon coupling term, $\lambda\phi^2T^2$, the action of the coupled scalar-tachyon fields model takes following form,
\begin{equation}
S=\int\dd^4x\sqrt{-g}[\frac{1}{16G}R-V(T)\sqrt{1-\ms^{-4}\partial_\mu T\partial^\mu T}-\frac{1}{2}\partial_\mu\phi\partial^\mu\phi-\frac{1}{2}m^2\phi^2-\lambda\phi^2T^2]~,\label{eq-cstaf}
\end{equation}
where $R$ is Ricci scalar and metric is taken as 
\eq{
\label{eq-metric}
\dd s^2=-\dd t^2+a^2(t) 
\Bigl(
    \frac{\dd r^2}{1-Kr^2}+r^2
    \bigl( \dd\theta^2+ \sin^2 \theta \dd\phi^2 \bigr)
\Bigr)~,\quad K=1~.
}
The second and third terms on the right hand of \refeq{Vphi} have been meshed into \refeq{cstaf}. The second term, the tension of D3-anti-D3 branes pair, has been considered as the tachyon's vacuum energy, $V_{0}$~, in \refeq{cstaf}~; And the third term, the contribution of the massive modes, have been replaced by the single tachyon field term in \refeq{cstaf}~.

In such unification, we can expect that this new model has following distinctive properties: 
\begin{enumerate}
\item When two branes are initially separated at a large distance, tachyon field is locked in the open string vacuum, $T\sim 0$ and $\dot{T}\sim 0$, by large value of $\phi$ through their coupling, $\lambda\phi^2T^2$. Meanwhile, universe inflates driven by the tension of branes like D-brane inflation; 
\item During the locked inflation, $\phi$ gets strongly redshifted, and eventually the coupling term fails to stabilize the system in the false vacuum. Therefore, tachyon condensation takes place, which indicates that the branes annihilates when their averaged distance becomes less than string length $M_s^{-1}$.
\end{enumerate}
Moreover, according to this model, universe is free of the big-bang singularity by undergoing a smooth bounce from a contracting phase to an expanding phase. The details are analyzed in \refsec{cem}~.  To sum up, a couple of technical remarks are warranted here: 
\begin{itemize}
\item This model applies a positive curvature since it is the only allowed choice to realize soft bouncing in the FLRW background without ghost. At the bouce point, the Hubble parameter vanishes, $H=0$. With the Friedmann equation, $H^2=-\frac{K}{a^2}+\frac{8\pi G}{3}\rho$, we obtain the scale factor at bouncing point, $a_{bounce}=\sqrt{\frac{3}{8 \pi G}}\sqrt{\frac{K}{\rho}}$. Clearly,  in absence of ghost, the energy density is positive, $\rho>0$. Therefore, $k$ should also be positive.

\item We model the D3-brane and anti-D3 brane to be able to pass each other when they collide,  with a certain probability for annihilation. Therefore, $\phi$ can take either positive or negative value. Moreover, after the tachyon condensation, D-brane and anti-D-brane become very thin but still exist. Thus the physical degree of freedom of $\phi$  does not disappear. And the effective action of $\phi$ is still valid after the tachyon condensation. 
\end{itemize}

\subsection{An Overview of Cosmic Evolution of the Coupled Scalar-Tachyon Model}
In this section, we offer an overview of the cosmic evolution driven by the coupled scalar-tachyon fields in the closed FLRW background, and sketch the energy density of fields and curvature, and the cyclic behavior of scale factor in  \refig{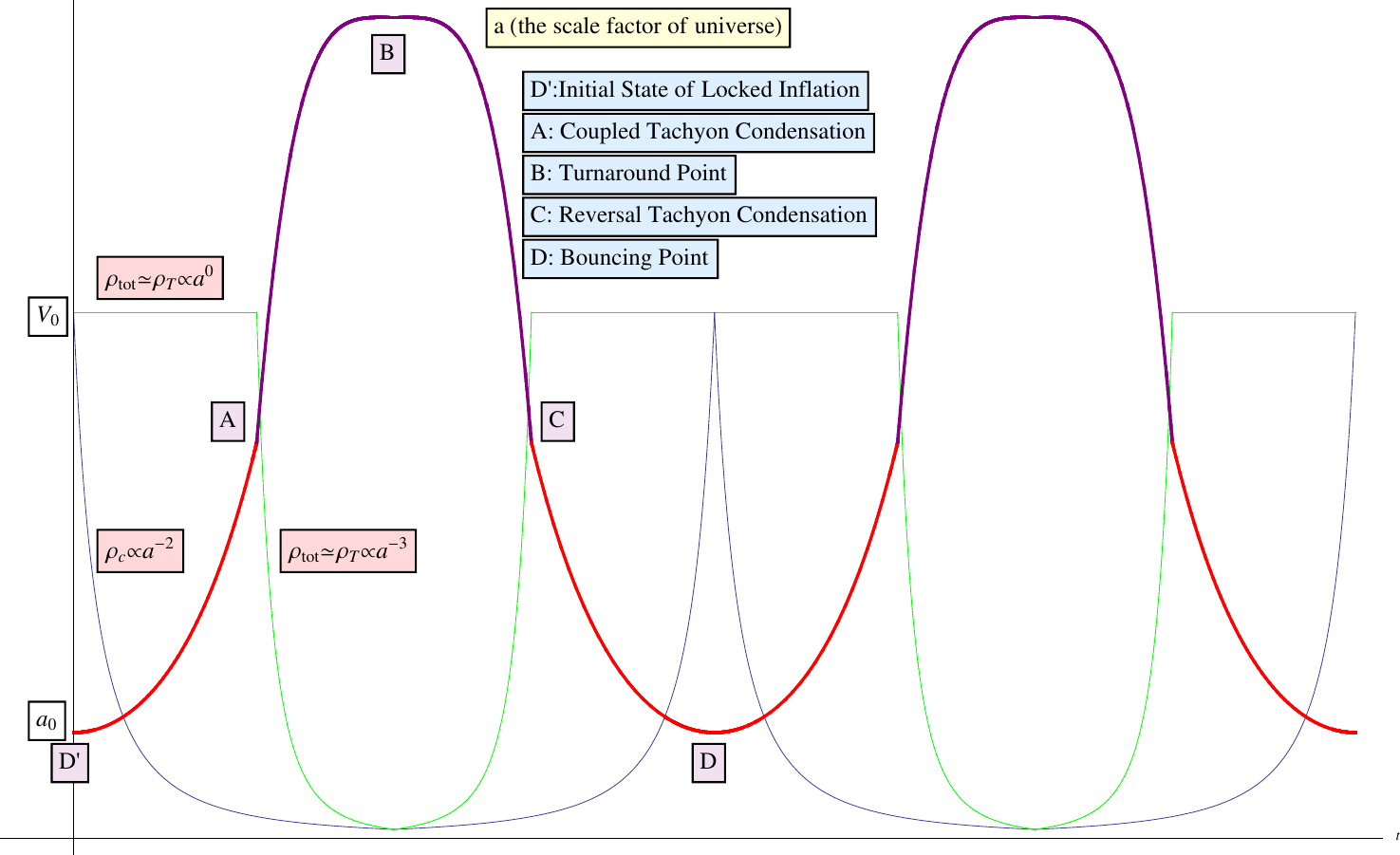}~. 

\fig{graphoverview.pdf}{The  sketches depicting  the  evolution of scale factor, total energy density of scalar field and tachyon field, and ``effective'' energy density of the curvature. The horizontal axis is time. The thick line (red and purple) shows the cyclic behavior of scale factor in pre-bounce, bounce, and post-bounce eras, and the red and purple colors are used to label $\ddot{a}>0$ (the dark energy dominated) era and $\ddot{a}<0$ (matter dominated) era respectively. The upper line (orange and green) is the total energy density of both fields. The orange and green color are used to label the dark energy dominated era and  matter dominated era respectively. The lower line (blue) is the ``effective'' energy density from the curvature. $D^\prime$ and $D$ are used to denote the initial state of locked inflation and bouncing point respectively, and the $A$ and $C$ are used to denote the Coupled Tachyon Condensation Transition and the Reversal of the Tachyon Condensation Transition respectively.}{0.9}{h!t} 

In \refig{graphoverview.pdf}, the cosmic evolution of scale factor, the total energy density of both fields, and the ``effective'' energy density of curvature are illustrated respectively. The thick lines (red and purple) indicate the cyclic behavior of scale factor with the pre-bounce, bounce, and post-bounce phases. The red and purple colors are used to label $\ddot{a}>0$ (the dark energy dominated) era and $\ddot{a}<0$ (matter dominated) era respectively. The upper line (orange and green) is the total energy density of both fields. The orange and green color are used to label the dark energy dominated era and  matter dominated era respectively. The lower line (blue) is the ``effective'' energy density of curvature. And \refig{graphoverview.pdf} is sketched from one minimum point of scale factor with $\dot{a}=0$ and $\ddot{a}>0$. In  \refig{graphoverview.pdf}, $D^\prime$ and $D$ are used to denote the initial state of locked inflation and bouncing point respectively.  The $A$ and $C$ are used to denote, respectively, the Coupled Tachyon Condensation  and the Reversal Tachyon Condensation.

Following the sketch map based on our model, we are giving a chronological list of each cosmic era of universe in one cycle, and briefly explain each of them at following. The detailed analyses of these cosmic eras are in consequential chapters.  

\begin{itemize}
\item {\bf Locked  inflation period ($D'\rightarrow A$ in \refig{graphoverview.pdf} ):}  

The initial state of this phase is naturally chosen at the minimum of scale factor, $\dot{a}=0$ and $\ddot{a}>0$. In this period, the D-brane and anti-D-brane are separated at large distance in extra dimension. The tension of them, $\rho_T= V_0$, behaves like dark energy, $w=-1$. During this phase, the energy density of tachyon field dominates, $\rho_T\propto a^0$ over the effective energy density of curvature, $\rho_c\propto a^{-2}$. Universe undergoes an exponential expansion until the D-anti-D-brane pair annihilating. Dynamically, the inflaton is locked in the false vacuum valley by the scalar-tachyon coupling term, therefore, we called this inflation period as ``locked inflation''.

\item {\bf Coupled tachyon condensation (around $A$ in \refig{graphoverview.pdf} ) :} 

Cosmologically, tachyon condensation can be viewed as a phase transition from dark energy to tachyon matter. During the locked inflation phase, the amplitude of scalar filed $\phi$, which is proportional to the distance $l$ between the D-anti-D-branes pair in extra dimension, gets strongly redshift so that the average distance between D-brane and anti-D-brane decreases. When $l$ decrease to the critical value $M_s^{-1}$, the D-brane and anti-D-brane start to annihilate efficiently. Therefore, the energy density of D-anti-D-brane pair releases into tachyon matter, {\it i.e.} $\omega_T=-1\longrightarrow \omega_T=0$~. And the locked inflation finishes. Universe evolves from the dark energy dominated era to a tachyon matter dominated era. 

One thing should be noticed that, incorporating with the scalar field, our model has a notable property: after coupled tachyon condensation, tachyon field acquires a non-zero effective vacuum and oscillates around it. This effective vacuum moves toward large $T$ in consequential expanding phase very slowly. And in a contracting phase, it would move back, and make ``Reversal Tachyon Condensation''  happen.  This novelty of coupled tachyon condensation in our model ensures that the universe would bounce from a contracting phase to an expanding phase as discussed in the consequential chapters.

\item {\bf Tachyon matter dominated period ($A\rightarrow B\rightarrow C$ in \refig{graphoverview.pdf} ):}

After the coupled tachyon condensation, universe is matter-dominated. Its evolution consists of following three phases:

{\bf The expanding phase ($A\rightarrow B$ in \refig{graphoverview.pdf} ):} During this phase, the energy density of tachyon matter is much larger than the effective energy density of curvature, $\rho_T\gg \rho_c$. With the initial condition of this phase, $\dot{a}>0$, universe undergoes a decelerating expansion, $a\propto t^{\frac{2}{3}}$, until approaching the turndown point.

{\bf The turndown point ( $B$ in \refig{graphoverview.pdf} ):} During the expansion, the effective energy density of curvature, $\rho_c\propto a^{-2}$, catches up the energy density of tachyon matter, $\rho_T\propto a^{-3}$ at the down point, $\dot{a}=0$ and $\ddot{a}<0$. After that, unverse evolve into a matter contracting phase.   

{\bf The contracting phase($B\rightarrow C$ in \refig{graphoverview.pdf} ):} In the contracting phase, the tachyon matter dominates again. With the initial condition of this phase, $\dot{a}<0$, universe undergoes a accelerating contraction, $a\propto t^{\frac{2}{3}}$, until the reversal tachyon condensation taking place.

\item {\bf The reversal tachyon condensation (around $C$ in \refig{graphoverview.pdf} )\bf :} 

The reversal tachyon condensation is a phase transition from tachyon matter to dark energy (tension of branes). In contracting phase, the scalar field, $\phi$, gets strong blue-shifted and pushes the effective vacuum of tachyon field back. Eventually, tachyon field would push up to the top of its potential with $T\sim 0$ and $\dot{T}\sim 0$~. Therefore, the energy density of tachyon matter is converted to the tension of D-anti-D-brane pairs. In D-brane scenario, it indicate the re-production of  D-anti-D-brane pairs by effective vacuum fluctuations. After the reversal tachyon condensation, universe is dark energy dominated again.

\item {\bf The bounce and the next cycle ($C \rightarrow D$ in \refig{graphoverview.pdf} ):}

After the reversal of the tachyon condensation, universe is dark energy dominated and undergoes an exponential contraction. During this contraction, the effective energy density of curvature, $\rho_c\propto a^{-2}$, catches up the energy density of tachyon matter, $\rho_T\propto a^{0}$, at the bouncing point, $\dot{a}=0$ and $\ddot{a}>0$. One may notice that this bouncing point is identical to the initial condition of the locked inflation, ($D'\sim D$ in \refig{graphoverview.pdf}). It implies that, after bouncing point, universe evolves into a new cycle of cosmological evolution same to what we discussed on above. 

\end{itemize}

To sum up this overview, in presence of the scalar field, tachyon condensation becomes reversible. And such reversal tachyon condensation,  a phase transition from tachyon matter'' to dark energy, is crucial for getting ``soft bounce'' solution in closed FLRW background.  In consequential chapters,  we are elucidating each stage listed above by the analytical calculation and the numerical simulation.

\secs{Cosmic Evolution of the Coupled Scalar-Tachyon Fields Model}
\label{sec-cem}
With the general action of coupled scalar-tachyon fields model, ~\refeq{cstaf}, the Friedman equation in the closed FRLW space-time becomes
\eq{H^2=-\frac{1}{a^2}+\frac{8\pi}{3\mpl^2}\Bigl(\frac{V(T)}{\sqrt{1-\ms^{-4}\dot{T}^2}}+(\frac{1}{2}m^2+\lambda T^2)\phi^2+\frac{1}{2}\dot\phi^2\Bigr),\label{eq-Friedmann}}
and the equations of motion for the tachyon and the scalar field are  
\eq{\frac{\ddot T}{\ms^4}+(1-\ms^{-4}\dot{T}^2)\biggl(\frac{3H\dot T}{\ms^4}+\frac{V'(T)+2\lambda\phi^2T\sqrt{1-\ms^{-4}\dot{T}^2}}{V(T)}\biggr)=0,
\label{eq-EoM-T}}
\eq{\ddot\phi+3H\dot\phi+(m^2+2\lambda T^2)\phi=0~.
\label{eq-EoM-phi}}
Armed with this equations, we are studying the evolution of the universe, starting with  the minimum $a(t)=a_0(t)$. To make our analysis clear, we sketch the potential of the coupled scalar-tachyon field with non-zero expectation value of $\phi$ field and denote each critical value on~\refig{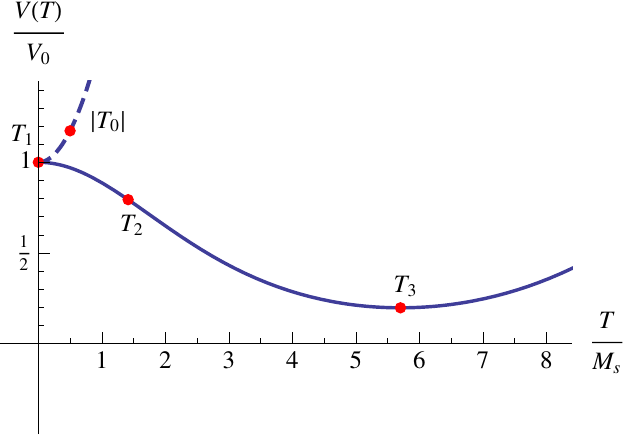}. 

\fig{Graph-Tn.pdf}{Timestamps of universe evolution presented in the potential of $T$. The dashed curve is the potential during locked inflation, and the solid one is after tachyon is released. The universe starts minimum at $T_0$. Locked inflation ends at $T_1$. Rolling inflation ends at $T_2$. Tachyon first reaches its \VEV{} at $T_3$.}{0.75}{h!t}

\refig{Graph-Tn.pdf} shows the expected stages in the figure of the potentials of $T$, with timestamps $(0, 1,...)$ in subscripts to indicate the time at which variables stand. We start the universe from its minimum, timestamped as $0$. Tachyon is initially locked at zero by the oscillating $\phi$. 
The bounce stage is short.  The curvature term is inflated away, and therefore the universe goes on to a locked inflation stage during which the amplitude of $\phi$ is also red-shifted.  When it drops to too low, it can no longer provide enough effective mass to $T$.   When it happens, $T$ starts to roll down. The bounce and locked inflation stages are indicated by  $0 \rightarrow 1$ in~\refig{Graph-Tn.pdf}, with  Point 1 being  the end of locked inflation.

At the beginning of tachyon's rolling,  when $\dot T^2\, \ll\, \ms^4$, there is another short period of inflation, {\it rolling inflation},  denoted by $1\rightarrow 2$ in \refig{Graph-Tn.pdf}.  During rolling inflation, $\ta{\phi^2}$ is damped and  $T$ picks up a vacuum expectation value.

Toward the end of the rolling inflation, the velocity of $T$ becomes large.   Inflation ends because the tachyon is matter-like. As shown in \refig{Graph-Tn.pdf}, tachyon is acceleratingly to catch up with its \VEV{} at this stage, indicated by $2\rightarrow 3$.

Once $T$ reaches its \VEV{}, it will oscillate and relax. Its kinetic energy is lost through expansion. It then follows that  the \VEV{}  of tachyon increases slowly before contraction happens. This happens after Point $T_3$ in \refig{Graph-Tn.pdf}.

Matter-like components have larger damping rates than the curvature term, so curvature gradually becomes important and eventually balances all positive energy contributions in the Friedman equation. The universe turns around at this point to enter a phase of contraction.  During contraction, $\phi$ is boosted and it drags $T$ back to zero. When the universe is again dominated by the vacuum energy of the tachyon, the positive energy behaves like a cosmological constant and stays constant for further contraction whereas the curvature energy increases at a rate of $a^{-2}$.  A bounce takes place when the curvature energy and vacuum energy of the tachyon reaches exact equality.  Even though the bouncing process is seemingly like the reverse of an expansion, the universe enters another cycle of expansion.

\ssecs{Locked Inflation Era}
We start following the universe evolution from the minimum of $a(t)$~. Tachyon is initially locked at its \VEV{} zero by the large amplitude of $\phi$. Curvature energy and vacuum energy of $T$ co-dominate and cancel each other at minimum $a$, but their equation of state difference still leads to an expansion  During the early co-dominant expansion stage, the curvature is red-shifted, and the value of expected vacuum of tachyon field becomes dominated. Since the co-dominant stage is short (with a time scale comparable to $\sqrt\frac{3\mpl^2}{8\pi V_0}$), universe evolves into locked inflation stage soon. 

The universe then goes on to the locked inflation, as indicated $0\rightarrow 1$ in \refig{Graph-Tn.pdf}. For non-zero value of $\phi$, the scalar-tachyon forms a false vacuum valley at $T=0$. 
During locked inflation, the interaction term produces a positive effective mass for $T$ with large amplitude of $\phi$. Therefore, the tachyon is locked in such false vacuum valley. And the scalar field $\phi$ is rolling along the false vacuum valley ($T=0$) with a  decreasing amplitude due to the inflation. In D-brane scenario, the locked inflation can be viewed as the two branes oscillate with a large amplitude, preventing the production of tachyon, i.e. $\lambda\ta{\phi^2}=\lambda V_0y^2\gg V_0/\ms^2$ and $\ta{T^2}\ll\ms^2$. The universe is dominated by $V_0$, the tension of the branes. Here $\ta{\ }$ means time averaging to remove oscillatory features.

 At this stage, the universe is dominated by the vacuum energy of $T$, 
 \begin{equation}
\rho_{fv}=V_0+\frac{1}{2}m^2\phi^2+\frac{1}{2}\dot{\phi}^2\add{\approx V_0}
\end{equation} 
where we suppose that the energy density of $\phi$ field is much smaller than the tension of D-anti-D-brane pair.  Therefore, the Hubble parameter is constant as long as $T$ is locked,
 \eq{\label{eq-H-li}
H^2=\frac{8\pi V_0}{3\mpl^2};
}
Since the tachyon field is very small and scalar field is relatively large, the scalar-tachyon coupling, $\lambda T^2\phi^2$, gives the tachyon field a positive effective mass square, $m_T^2\equiv2\lambda \phi^2-\frac{V_0}{2M_s^2}$~, in its equation of motion, 
\begin{equation}  
\label{ddTr}
\frac{1}{M_s^4}\ddot{T}+(-\frac{1}{2M_s^2}+\frac{2\lambda\phi^2}{V_0} )T=0~.
\end{equation}
So the tachyon field is kept from rolling down at $\ta{T^2}\ll\ms^2$ as long as $m_T^2>0$. Moreover, the equations of motion for the scalar field $\phi$ is also simplified to be: 
\begin{equation}  \label{ddphir}
\ddot{\phi}+3H\dot{\phi}+m^2\phi=0~.
\end{equation}
accordingly. Henceforth, the universe expands exponentially from~\refeq{H-li} driven by vacuum energy,
\eq{
a \propto e^{\sqrt\frac{8\pi V_0}{3\mpl^2}\,t}~,\label{eq-loin}
} 
and $\phi$ keeps oscillating and gets redshifted with
\eq{
\phi\propto a^{-\frac{3}{2}}e^{i\sqrt{m^2-\frac{9}{4}H^2}\,t}~.
}
From the oscillation term we notice $\phi$ may pass zero many times during oscillation. The physical interpretation is that a  D3-brane and an anti-D3-brane can  meet many times before total annihilation. This is because the maximum distance between the branes (i.e. the amplitude of oscillation) is large. When they meet, this leads to a large relative velocity between the branes so time is insufficient for annihilation. 
One the other hand, $a^{-\frac{3}{2}}$, tells us that the scalar field get strongly redshifted. In string theory, it means the maximum distance between the branes are decreasing. When the maximum distance is small enough, the two branes will annihilate,  putting an end to locked inflation. The proposal of locked inflation here  is analogy to the New Old Inflation model of  Dvali and Kachru~\cite{Dvali:2003vv}, but this one is string-inspired.  We will also  go further by investigating two different ending conditions of locked inflation.

The one proposal for ending of locked inflation is caused by the change of the sign of $m_T^2$, the effective mass square of tachyon, marking the beginning of the tachyon's rolling down its potential hill. 
Given $m_T^2\, = 2\, \lambda\ta{\phi^2}\, -\, V_0/2\, \ms^2$, the critical $\ta{\phi^2}$ at the ending of locked inflation, $\ta{\phi_c^2}$, is derived from
\eq{ \label{eq-phic}
m_T^2=0\Rightarrow\ta{\phi^2}=\ta{\phi_c^2}\equiv V_0/4\lambda\ms^2~.
}
Therefore when $\phi^2$ reaches an expectation value of $\phi_c^2$, it can no longer hold the tachyon at the top of its potential hill. And $T$ starts  to roll down, ending the locked inflation.

There exists another way to exit locked inflation.  When $\phi$  oscillates much slower than tachyon, in which case as $\phi$ passes zero, tachyon will have sufficient chance/time to roll down. 
Therefore even if the effective mass square of $\phi$ is lower than $V_0/2\ms^2$, locked inflation can still hold if the amplitude of $\phi$ is large enough: the larger amplitude of $\phi$, the shorter the time $\phi$ it stays in the region $\ta{\phi^2}<\ta{\phi_c^2}$. 
We'll call this ending condition ``mass ending'' and denote  it by
\begin{equation}
m_\phi^2 \, \ta{\phi^2} =\frac{V_0}{2\ms^2}\ta{\phi_c^2},
\end{equation}
where $m_\phi$ is the effective mass of $\phi$.

These two conditions are connected at $m_\phi^2=V_0/2\ms^2$. When $m_\phi^2$ is larger, then the normal ending discussed earlier should be adopted, and a smaller value of  $m_\phi^2$ calls for ``mass ending.'' We can see from the mass ending condition that  a lower $m_\phi^2$ makes $\ta{\phi^2}$ larger at the end of locked inflation.  This implies an earlier ending of the locked inflation than previously thought of. Although the two endings have different conditions, they have to meet  the same criterion -- the sufficiency of time for brane annihilation.

Although the expectation value of tachyon is zero, $\ta{T}=0$, $T$ still has a nonzero fluctuation $\ta{T^2}\ne 0$. The back-reaction of it may be large by generating an additional effective mass square $2\lambda\ta{T^2}$ for $\phi$. In cases where the dominant part of effective mass of $\phi$ differs, the emergence of those two fields is also different. Therefore, separate consideration is needed for those cases, specified by where $m^2/\lambda$ is inserted into the relation of $\ta{T_1^2}<\ta{T_0^2}<\ta{T_2^2}\sim\ms^2$. For the same reason, two ending conditions of locked inflation are also considered separately. In all, there are $2\times 3=6$ cases as labelled $a\rightarrow f$ in \reftab{Labels of the six cases for separate consideration}.

\tab{Labels of the six cases for separate consideration}{
\vspace{0.15in}
\begin{tabular}{|c||c|c|c|}\hline
&$2\lambda\ta{T_2^2}>m^2>2\lambda\ta{T_0^2}$&$m^2<2\lambda\ta{T_1^2}$&$2\lambda\ta{T_0^2}>m^2>2\lambda\ta{T_1^2}$\\\hline\hline
Normal Ending&$a$&$b$&$c$\\\hline
Mass Ending&$d$&$e$&$f$\\\hline
\end{tabular}
}

First we consider case $a$, in which $2\lambda\ta{T_2^2}>m^2>2\lambda\ta{T_0^2}$ with normal ending. In this case, $m^2>2\lambda\ta{T^2}$ holds throughout the entire period of  locked inflation. The constant mass gives $\ta{\phi^2}\propto a^{-3}$. Tachyon however has a varying mass $\lambda\ta{\phi}^2\propto a^{-3}$, so the solution of its equation of motion is $\ta{T^2}\propto a^{-3/2}$. Such relations implies that  the normal ending of locked inflation occurs at:
\eq{\ta{\phi_1^2}=e^{-3N_L}\ta{\phi_0^2}=\ta{\phi_c^2},}
where $\phi_c$ is defined in~\refeq{phic}. So the expected e-folds of locked inflation for case $a$ is
\eq{N_L^{(a)}=\frac{1}{3}\ln\frac{4\lambda\ms^2\ta{\phi_0^2}}{V_0}.}
Also,
\eq{\ta{T_1^{(a)2}}=\ta{T_0^2}\sqrt{\frac{V_0}{4\lambda\ms^2\ta{\phi_0^2}}}.}

Similarly, we can analyze  other cases of interest. For case $b$, because $2\lambda\ta{T^2}$ dominates $\phi$'s mass, there are $\ta{\phi^2}\propto a^{-2}$ and also $\ta{T^2}\propto a^{-2}$. Case $c$ has a transition point at $2\lambda\ta{T^2}=m^2$, before which we apply the analysis for case  $b$ and after that the analysis for case $a$ applies. The two sections are then connected at the transition point to obtain the complete  solution.   Cases $d,e,f$ are more or less similar to $a,b,c$, except that the ending condition should be replaced by $m_\phi^2\ta{\phi_1^2}=\frac{V_0}{2\ms^2}\ta{\phi_c^2}$. So we get the six scenarios as listed 
in~\reftab{Locked inflation of six separate cases}.

The six cases then correspond to different regions in the moduli space, dictated by the required conditions in the corresponding cases. 
For example, $a$ requires $2\lambda\ta{T_2^2}>m^2>2\lambda\ta{T_0^2}$ as in \reftab{Labels of the six cases for separate consideration}. All such restrictions of moduli space are collected, analyzed and presented in~\refsec{Analysis}.   We can  temporarily ignore these restrictions when we  follow  the evolution of the universe.

\tab{Locked inflation of six separate cases}{%
\vspace{0.15in}
\begin{tabular}{|c||c|c|c|}\hline
&$N_L$&$\ta{T_1^2}$&$\ta{\phi_1^2}$\\\hline\hline
$a$&$\displaystyle\frac{1}{3}\ln\frac{4\lambda\ms^2\ta{\phi_0^2}}{V_0}$&$\displaystyle\ta{T_0^2}\sqrt\frac{V_0}{4\lambda\ms^2\ta{\phi_0^2}}$&$\displaystyle\frac{V_0}{4\lambda\ms^2}$\\\hline
$b$&$\displaystyle\frac{1}{2}\ln\frac{4\lambda\ms^2\ta{\phi_0^2}}{V_0}$&$\displaystyle\frac{V_0}{4\lambda\ms^2\ta{\phi_0^2}}\ta{T_0^2}$&$\displaystyle\frac{V_0}{4\lambda\ms^2}$\\\hline
$c$&$\displaystyle\frac{1}{6}\ln\frac{32\lambda^3\ms^4\ta{\phi_0^2}^2\ta{T_0^2}}{m^2V_0^2}$&$\displaystyle\sqrt\frac{m^2V_0\ta{T_0^2}}{8\lambda^2\ms^2\ta{\phi_0^2}}$&$\displaystyle\frac{V_0}{4\lambda\ms^2}$\\\hline
$d$&$\displaystyle\frac{1}{3}\ln\frac{8\lambda m^2\ms^4\ta{\phi_0^2}}{V_0^2}$&$\displaystyle\ta{T_0^2}\sqrt\frac{V_0^2}{8\lambda m^2\ms^4\ta{\phi_0^2}}$&$\displaystyle\frac{V_0^2}{8\lambda m^2\ms^4}$\\\hline
$e$&$\displaystyle\frac{1}{4}\ln\frac{16\lambda^2\ms^4\ta{\phi_0^2}\ta{T_0^2}}{V_0^2}$&$\displaystyle\sqrt\frac{V_0^2\ta{T_0^2}}{16\lambda^2\ms^4\ta{\phi_0^2}}$&$\displaystyle\sqrt\frac{V_0^2\ta{\phi_0^2}}{16\lambda^2\ms^4\ta{T_0^2}}$\\\hline
$f$&$\displaystyle\frac{1}{6}\ln\frac{128\lambda^3m^2\ms^8\ta{\phi_0^2}^2\ta{T_0^2}}{V_0^4}$&$\displaystyle\sqrt\frac{V_0^2\ta{T_0^2}}{16\lambda^2\ms^4\ta{\phi_0^2}}$&$\displaystyle\frac{V_0^2}{8\lambda m^2\ms^4}$\\\hline
\end{tabular}
}

\subsection{Coupled Tachyon Condensation Transition}
\label{sec-ctc}
Being different from the tachyon condensation in single tachyon field model, the coupled tachyon condensation of this coupled scalar-tachyon field model involves the scalar field and the coupling term between scalar and tachyon field. The coupled tachyon condensation preserves the feature of annihilation of D-anti-D-branes pairs that, the tension of branes decay into the tachyon matter and these tachyon matter condensates after branes annihilating.  In the single tachyon field models, the tachyon field would roll toward infinity after tachyon condensation which renders the tachyon condensation to be irreversible. The novelty of the coupled tachyon condensation however are that tachyon field relaxes around the ``effective vacuum'' of the coupled potential rather than roll toward the infinity after such condensation. The position of the ``effective vacuum'' of the coupled potential is approximately determined by the value of the amplitude of the $\phi$ field. The larger amplitude of $\phi$ implies the ``effective vacuum'' being closer to the origin, $T=0$. Therefore, the coupled tachyon condensation provides a possible way to achieve a reversal  of the tachyon condensation. Specifically, in our model, such ``reverse tachyon condensation'' is realized by the increasing amplitude of $\phi$ during contraction phase, see~\refsec{rtc}. In this section, we studied the coupled tachyon condensation in detail by doing analytical analysis and numerical simulation respectively.

Generally, the coupled tachyon condensation consists of three main parts, tachyon rolling which $\dot{T}$ evolves from $0$ to $\sqrt{\frac{2}{3}}M_s$, tachyon accelerating $\dot{T}\rightarrow M_s$, and tachyon field oscillation around effective vacuum.  

\paragraph{{\bf Tachyon Rolling and its Inflation}}
When $\phi$ can no longer lock $T$ at its origin, $T$ will start to roll down to either side of the potential hill.  At the beginning when $T$ is small and thus 
$\dot T^2\ll\ms^4$, the \VEV{} of tachyon still dominates the universe. So this stage (indicated by $1\rightarrow 2$ in~\refig{Graph-Tn.pdf}) is an additional  epoch of inflation which is usually much shorter than locked inflation. Although most two-field models neglect this stage,  we consider it here because it causes a delay between the motion of $T$ and its \VEV{}. Such delay determines the length of following stages, and therefore should not be neglected in our analysis.

To good approximation we can set the mass squared of $T$ to  $-\ms^2/2$, whereas the effective mass from interaction is neglected. Generically  this approximation is  valid only when the number of e-folds of rolling inflation $N_R>1$. It  is also applicable to our case   even if $N_R<1$. The rolling inflation stage is then short and  itself can thus  be neglected for e-folding  counting.  
The equation of motion of $T$ then simplifies to
\eq{\ddot T+3H\dot T-\frac{1}{2}\ms^2T=0,}
with a solution of
\eq{T=T_1e^{(\frac{1}{\sqrt{2}}\ms-\frac{3}{2}H)(t-t_1)}.}
So the e-folds of rolling inflation is
\eq{N_R=H(t_2-t_1)=\frac{1}{\frac{\ms}{\sqrt{2}H}-\frac{3}{2}}\ln\frac{T_2}{T_1},}
where we approximately choose the end of rolling inflation to be at $T_2=\sqrt{2}\ms$ for simplicity of future calculations. From this equation we can see it is difficult to get a large $N_R$ because $H\ll\ms$.

Tachyon would later oscillate about its \VEV{} after rolling inflation, so keeping track of how the tachyon \VEV{} moves is  necessary. 
We define $T_V$, the \VEV{} of $T$ as
\eq{\left.\frac{\partial\rho}{\partial T}\right|_{T=T_V}=0.}
and 
\eq{\label{eq-TV}
T_Ve^\frac{T_V}{\sqrt{2}\ms}=\frac{V_0}{2\sqrt{2}\lambda\ms\ta{\phi^2}\sqrt{1-\dot T^2/\ms^4}}.
}
It is important to note the $\dot T$ in the above equation is the real time velocity of tachyon, not the velocity  of $T_V$. At point $2$, the equation becomes
\eq{\label{eq-TV2}
T_{V2}e^\frac{T_{V2}}{\sqrt{2}\ms}=\frac{V_0}{2\sqrt{2}\lambda\ms\ta{\phi_2^2}\sqrt{1-\dot T_2^2/\ms^4}},
}
in which at $T_2=\sqrt{2}\ms$ there is $1-\dot T_2^2/\ms^4\sim<1$.

To get $T_{V2}$, we still need to construct the relation between $\ta{\phi_2^2}$ and $\ta{\phi_1^2}$. The evolution of $\phi$ however varies under two different situations -- $m^2<2\lambda T^2$ through rolling inflation, and a transition to $m^2>2\lambda T^2$ occurs during rolling inflation. Here we don't consider the case with $m^2>2\lambda T_2^2$, because then the scalar field is too massive. For demonstration, such huge mass would not give any distinctive effect either. 

First for cases $b,e$, there is always $m^2<2\lambda T^2$. Therefore
\eq{\ta{\phi_2^{(b,e)2}}=\ta{\phi_1^2}e^{-3N_R}\sqrt\frac{\ta{T_1^2}}{T_2^2}=\ta{\phi_1^2}e^{-(\frac{3}{2}+\frac{\ms}{\sqrt{2}H})N_R}.}

The second possibility is that there exists a transition to $2\lambda T^2>m^2$ at $T^2=m^2/2\lambda$, for the cases of $a,c,d,f$. Suppose the transition happens at $T_\alpha$, i.e. $m^2=2\lambda T_\alpha^2$, combining the above two processes gives
\eq{\ta{\phi_2^{(a,c,d,f)2}}=\ta{\phi_1^2}\frac{me^{-3N_R}}{\sqrt{2\lambda}T_2}.}

We can see from these results the relationship is actually $\ta{\phi^2}\propto a^{-3}m_\phi^{-1}$, and the physics in it is quite clear. $\ta{\phi^2}$ is proportional to $a^{-3}$ because $\phi$ is a fast rolling field, so its amplitude is dampened by universe expansion by $a^{-\frac{3}{2}}$. It is then inversely proportional to $m_\phi$ because the kinetic energy is transferred to more potential energy as $m_\phi$ gets smaller. We can also get this result from the energy aspect of view. The potential energy of $\phi$ is proportional to $m_\phi^2$, however $\phi$ is a harmonic oscillator, so potential energy is only half of its total energy when averaged with time, from which we can infer its total energy is proportional to $m_\phi$. Its total energy can be represented as $m_\phi^2\ta{\phi^2}\propto m_\phi$, and  we get $\ta{\phi^2}\propto m_\phi^{-1}$.

\paragraph{\bf Accelerating Tachyon}
The period indicated by $2\rightarrow 3$ is an accelerating process for tachyon with its velocity $\dot T\rightarrow\ms^2$. When $T$ catches up with $T_V$, acceleration stops, signalling the end of this stage.   The length of this stage  hinges on the number of e-foldings from rolling period, for reasons explained above.  

If we restrict the moduli space inside $m^2<2\lambda\ms^2$, there is then $\ta{\phi^2}\propto a^{-3}T^{-1}$. To calculate the speed of $T_V$, we differentiate~\refeq{TV} w.r.t. time and get
\eq{\Bigl(\frac{1}{T_V}+\frac{1}{\sqrt{2}\ms}\Bigr)\dot T_V=3H\Bigl(1-\frac{\dot T^2}{\ms^4}\Bigr)+\frac{\dot T}{\sqrt{2}\ms}\biggl(1+\frac{\sqrt{2}\ms}{T}-\frac{T}{T_V}e^\frac{T-T_V}{\sqrt{2}\ms}\biggr).}
Neglecting $3H(1-\dot T^2/\ms^4)$, we have the solution of
\eq{\frac{T_V}{T}e^\frac{T_V-T}{\sqrt{2}\ms}+\frac{T}{\sqrt{2}\ms}=C,\label{eq-TV3}}
where $C$ is a constant of  integration.

Choosing point 2 as $T_2=\sqrt{2}\ms$ and point 3 as $T_{V3}=T_3$, by which we mean at point 3 the tachyon field reaches its \VEV{}  and starts to decelerate, we apply the solution~\refeq{TV3} to the start and end points of this stage and arrive at
\eq{T_3=T_{V3}=T_{V2}e^{\frac{T_{V2}}{\sqrt{2}\ms}-1}.}

The universe is matter-like during $2\rightarrow 3$.  At the beginning, tachyon dominates with equation of state $\omega_T\approx 0$. Later  when $\lambda T^2\ta{\phi^2}$ comes to dominate, $T$ is already large and its variation is negligible so $\lambda T^2\ta{\phi^2}\propto Ta^{-3}\propto a^{-3}$. Since both components are matter-like, we can  also compute  the e-folds of this stage
\eq{N=\frac{2}{3}\ln\frac{3H_2T_3}{2\ms^2}.}
With a bit of calculation, we can see $N=2N_R+ \ldots $ where $\ldots$ indicate other contributions. Such dependence on $N_R$ is exactly as expected. The longer rolling inflation lasts, the larger delay is between $T$ and $T_V$, and thus the more e-folds are required in order for $T$ to catch up with $T_V$.

\paragraph{\bf Tachyon Field Oscillation and its Amplitude}
When $T$ reaches its \VEV{} $T_V$ for the first time, it starts to oscillate around its \VEV{}. We are showing these properties in detail and compute the amplitude of such oscillation now. 

Define $x\equiv 1-\dot T^2/\ms^4$. The equation of motion of $T$ can  then be  transformed to
\eq{\frac{\dot x}{x} = 2 \dot T \biggl(\frac{3H\dot T}{\ms^4}-\frac{1}{\sqrt{2}\ms}+\frac{4\lambda\ta{\phi^2}Te^{T/\sqrt{2}\ms}}{V_0}\sqrt{x}\biggr).}
Here we neglect $3H\dot T/\ms^4$ term because it is much smaller than the other two. Adopting~\refeq{TV} at $T=T_V$ and $\ta{\phi^2}\propto a^{-3}T^{-1}$ to eliminate $\ta{\phi^2}$, we come to the following equation
\eq{\frac{\dd x}{2x}=\frac{\dd T}{\sqrt{2}\ms}\biggl(\frac{a_r^3}{a^3}e^\frac{T-T_r}{\sqrt{2}\ms}\sqrt\frac{x}{x_r}-1\biggr),}
where subscript $r$ represents the time when the last \VEV{} of $T$ was reached, or when $T$ reaches its \VEV{} next time. The solution is
\begin{equation}\frac{T-T_r}{\sqrt{2}\ms}=\ln\sqrt\frac{x_r}{x}-\ln\Bigl(1-\frac{T-T_r}{\sqrt{2}\ms}\Bigr)+3N_r\label{eq-ato}
\end{equation}
where $N_r\equiv\ln a/a_r$.

From this we can calculate the amplitude of $T$. Because  $\dot T^2\rightarrow\ms^4$ at $T=T_V$, we have $x_r\ll 1$. The amplitude adopted here are the two fastest distances $T$ can reach from $T=T_V$ (left and right) during one oscillation cycle, where $\dot{T}=0$ that $x=1$.  $T$ has an asymmetric potential w.r.t. $T_r$, so its amplitude on the sides of $T>T_r$ and $T<T_r$ are different. Defining the amplitudes at $T>T_r$ and $T<T_r$ as $T_+$ and $T_-$ respectively, we now calculate them in turn. 

First for $T_+$ when $T>T_r$,  the~\refeq{ato} is simplified to
\eq{T_+=T_r+\sqrt{2}\ms(1-e^{3N_r}\sqrt{x_r})\approx T_r+\sqrt{2}\ms. \label{eq-roa}}
It is because of, for  $x_r\ll 1$, the term $\ln\sqrt x_r$ has a large negative value and should be almost cancelled by the term $\ln\Bigl(1-\frac{T-T_r}{\sqrt{2}\ms}\Bigr)$ as $1-\frac{T-T_r}{\sqrt{2}\ms}\rightarrow 0$ by leaving a relatively small residue $\frac{T-T_r}{\sqrt{2}\ms}-3N_r$. Furthermore, since $\ln\Bigl(1-\frac{T-T_r}{\sqrt{2}\ms}\Bigr)$ is very steep when $1-\frac{T-T_r}{\sqrt{2}\ms}\rightarrow 0$, the $T$ turn back swiftly as it reaches $T_+$ and the plot of $T$ should be sharp at $T=T_+$. Our numerical simulation well confirms this point, see \refig{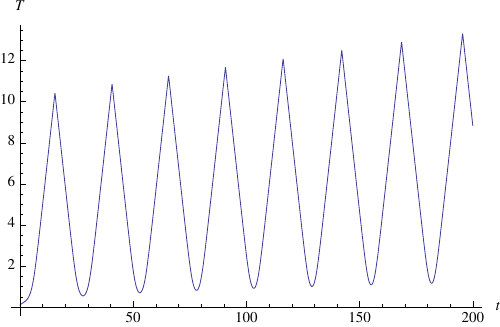}, where the plot of tachyon field have a sharp peak in each oscillation.

Similarly, for $T_-$ when $T<T_r$,  
we have
\eq{T_-=T_r+(\ln\sqrt{x_r}+3N_r)\sqrt{2}\ms\approx T_r+\sqrt{2}\ms\ln\sqrt{x_r}. 
\label{eq-loa}} 
Because $x_r\ll 1$, $T$ can go far away from $T_r$ when $T<T_r$, so that the  amplitude on $T<T_r$ is larger than the amplitude on $T>T_r$ of tachyon field oscillation. Furthermore, no term of the~\refeq{ato} is steep in the region $T<T_r$,  so the $T$ should approach and leave from $x=1$ on $T<T_r$ smoothly,  i.e. it turns back from $T<T_r$ smoothly.

In summary, the amplitude of tachyon field oscillation are $\sqrt{2}\ms$ on $T<T_r$  and  $\sqrt{2}\ms\ln\sqrt{x_r}$ on $T>T_r$ respectively. And the tachyon field turns back from the region $T>T_r$ sharply and from the region $T<T_r$  smoothly. These two signatures are well confirmed in the~\refig{sTe.pdf}, a plot of our numerical simulation results.

Now we present a numerical simulation for the coupled tachyon condensation to confirm our analytical analysis on above.  After the locked inflation, the equations of motion for the coupled scalar-tachyon fields model, \refeq{Friedmann}$-$\refeq{EoM-phi}, can be much simplified for numerical simulation as shown in~\refsec{nsa}~. Utilizing the~\refeq{sddTe}, we perform the numerical simulation of the coupled tachyon condensation as following. $t=0$ is chosen at the ending of the locked inflation phase. The initial conditions are taken as: $T[0]=0.1$, $\dot{T}[0]=0.1$, $L=\frac{2}{3}*10^3$, $k=0.1$~. And we also take the convention, $\ms=10^{-3}M_p=1$, in this numerical simulation. The tachyon field, the Equation of State and energy density of tachyon field in coupled tachyon condensation are plotted in~\refig{sTe.pdf}, \refig{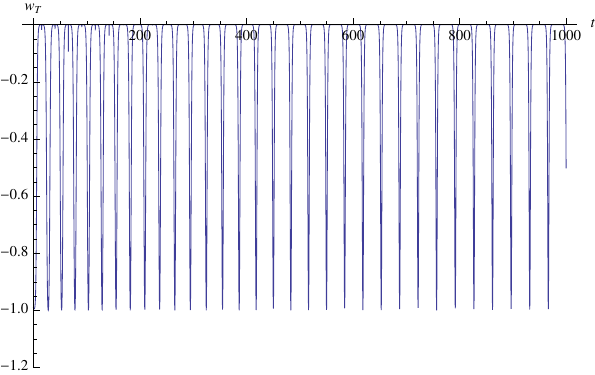},and \refig{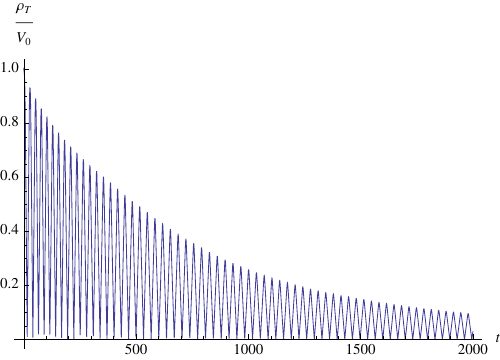} respectively. 
 
\begin{enumerate}

\item In~\refig{sTe.pdf}, tachyon field is plotted from $t=0$ to $t=200$~. On the left side of the effective vacuum, $T_v$ where $\ddot{T}=0$, the term $-\frac{1}{\sqrt{2}}\tanh(\frac{T}{\sqrt{2}})$ dominates, and on the right side of the $T_v$, the term $k\frac{L^2}{(t+L)^2}\sqrt{1-\dot{T}^2}\cosh(\frac{T}{\sqrt{2}})$ dominates, see ~\refeq{sddTe}.  Once the tachyon field rolling down from the top of its potential, the term $-\frac{1}{\sqrt{2}}\tanh(\frac{T}{\sqrt{2}})$ accelerate the tachyon field, $\dot{T}\rightarrow1$. After the tachyon field passing the $T_v$, the term $k\frac{L^2}{(t+L)^2}
\sqrt{1-\dot{T}^2}\cosh(\frac{T}{\sqrt{2}})$ becomes dominated. This term decelerates the rolling of tachyon field, $\dot{T}\rightarrow-1$, and finally makes the tachyon field roll back and pass the $T_v$ again. Such processes take place repeatedly, so the evolution of tachyon field plotted in~\refig{sTe.pdf} possess an oscillatory feature. 

\fig{sTe.pdf}{The tachyon field is plotted in the coupled tachyon condensation process from $t=0$ to $t=200$. The expectation value of tachyon field, $T_v$, gradually increases with time, and the tachyon field oscillates around such expectation value swiftly.}{0.70}{h!t}

\item The Equation of State of tachyon field, $\omega_T=-(1-\dot{T}^2)$, is plotted from $t=0$ to $t=1000$ in~\refig{swTe.pdf}. According to ~\refeq{sddTe}, the effective force terms, $k\frac{L^2}{(t+L)^2}\sqrt{1-\dot{T}^2}\cosh(\frac{T}{\sqrt{2}})$ and $-\frac{1}{\sqrt{2}}\tanh(\frac{T}{\sqrt{2}})$, are suppressed by the factor $1-\dot{T}^2$ when $|\dot{T}|\rightarrow1$. Therefore, the time interval, which $|\dot{T}|$ staying around $1$, should be much longer than that around 0. Henceforth, even through $\omega_T$ occasionally reaches $-1$,  the average value of $\omega_T$ also should be almost $0$. Our numerical simulation confirms such argument. In the~\refig{swTe.pdf}, we see that the $\omega_T$ evolves from the initial value $\omega_T\simeq -1$ to $\omega_T\simeq 0$ swiftly, which correspond to the process of tachyon rolling down. Moreover, the time interval that $\omega_T$ staying around $0$ becomes longer and longer in time. It indicate that, during the coupled tachyon condensation, the energy density of D-anti-D-branes pair, which is dark energy-like, releases into tachyon matter.

\fig{swTe.pdf}{The evolution of the Equation of State of tachyon field $\omega_T$ from $t=0$ to $t=1000$. The equation of state of tachyon field  evolves into the later region $\omega_T\simeq 0$ from the initial state $\omega_T\simeq -1$ swiftly. In the later region, even though it occasionally reaches $-1$, it generally stays around $0$. So the average value of $\omega_T$ is almost $0$ in later region, which satisfies the requirement of tachyon condensation $\omega_T\rightarrow 0$.}{0.70}{h!t}

\item In~\refig{srhoTe.pdf}, the energy density of tachyon field $\rho_T=V(T)(1-\dot{T}^2)^{-\frac{1}{2}}$ is plotted from $t=0$ to $t=2000$. Despite the fast oscillation, the averaged value of the energy density is decreasing with time, and $\rho_T\propto a^{-3}$ approximately. It is consistent to the expectation that, after the coupled tachyon condensation, the tachyon field behaves like normal matter with $\omega_T\sim 0$, and get redshifted during the consequential expansion phase.   

\fig{srhoTe.pdf}{The evolution of the energy density of tachyon field is plotted from $t=0$ to $t=2000$. And it decreases as $\rho_T\propto a^{-3}$ in this process. }{0.70}{h!t}

\end{enumerate}

To sum up the numerical simulation of the coupled tachyon condensation, we emphasize several additional features of these numerical results. For given $T$, the term $k\frac{L^2}{(t+L)^2}\sqrt{1-\dot{T}^2}\cosh(\frac{T}{\sqrt{2}})$ gradually decreases with time due to the red-shift of the $\phi$. Therefore, the effective vacuum, $T_v$, increases gradually, see \refig{sTe.pdf}. On the other hand, the value of $T_v$ would be larger than $\sqrt{2}$ soon, which indicates that the annihilation process of D-anti-D-branes pair and the coupled tachyon condensation are completed.  Furthermore, the two distinctive features, the sharp peaks and smooth troughs as shown in~\refig{sTe.pdf}, are consistent to our previous analytical analysis, see~\refeq{roa}, ~\refeq{loa} and their explanation.  

\ssecs{Tachyon Matter Domination Era: Expansion, Turnaround, and Contraction}
\label{sec-turnaround}

After the coupled tachyon condensation, the energy density of tension of D-anti-D branes pairs releases into the tachyon matter.  Universe evolves into a tachyon matter dominated era, and undergoes a matter-dominated decelerating expansion,
\begin{equation}
a=a_{ec}[\frac{3}{2}\frac{\sqrt{V_0}}{\tilde{M_p}}(t+\frac{2}{3}\frac{\tilde{Mp}}{\sqrt{V_0}})]^\frac{2}{3}~,
\end{equation}
with the initial conditions, $\dot{a}_{ec}>0$, where we use the subscript $_{ec}$ to label the quantities at the end of coupled tachyon condensation phase. 

After the locked inflation and coupled tachyon condensation stages, the energy density of the curvature term, $\rho_c$, takes very small fraction in the total energy bill and can be neglected. However, during the tachyon matter dominated expansion phase, the density density of curvature, $\rho_c\propto a^{-2}$, is catching up the energy density of tachyon matter, $\rho_T\propto a^{-3}$, gradually. Eventually, universe reaches a turnaround point at where the energy density of curvature and tachyon matter cancel each other.  At the same time, the scale factor also reaches its maximum, $a_{max}=V_0 a_{ec}^3\tilde{M}_p^{-2}$ and $\ddot{a}<0$. 

After the turnaround point, universe start to contract. During this contraction,  tachyon matter dominates over the curvature again. Therefore, universe undergoes a tachyon matter dominated contraction,
\begin{equation}
a=a_{ec}[\frac{2}{3}\frac{\sqrt{V_0}}{\tilde{M_p}}(-t+\frac{2}{3}\frac{\tilde{M_p}}{\sqrt{V_0}}\frac{a_{max}^3}{a_{ec}^3})]^\frac{2}{3}~, \label{eq-mdc}
\end{equation}
until the Reversal Tachyon Condensation(discussed in detail at next section) taking place. 
If there is no Reversal Tachyon Condensation, which ensures the tension of D-anti-D branes pairs dominate again after tachyon matter dominated contraction, this contraction, \refeq{mdc}, would cause a cosmological singularity at $t=\frac{2}{3}\frac{\tilde{M_p}}{\sqrt{V_0}}\frac{a_{max}^3}{a_{ec}^3}$~. 
 
Now we turn our attention to the dynamics of background fields during these tachyon dominated eras. And we are discussing how the Reversal Tachyon Condensation start up after tachyon matter dominated contraction.    

As an auxiliary and sub-dominated background field, $\phi$ gets redshift and blueshift in the tachyon matter dominated expanding phase and contracting phase respectively. According to \refeq{EoM-T}~, the expected vacuum, $T_v$ for $\ddot{T}=0$, is determined by the amplitude of $\phi$. During the expanding phase, the amplitude of $\phi$ decreases, and $T_v$ runs away from the hill of tachyon field potential, see~ \refig{sTe.pdf} for instance. On other hand, during the contracting phase, the amplitude of $\phi$ increases and pushes the tachyon back to the hill of its potential through the scalar-tachyon coupling term. Qualitatively, according to~\refeq{sddT}~, the expectation value of tachyon field  is proprotional to the E-folding number of universe in the tachyon matter dominated epoch, $T_v\propto 3N_p$~.

Once the tachyon field is pushed back to the top of its potential hill with $T\sim 0$ and $\dot{T}\sim 0$ after the tachyon matter dominated contraction, the energy density of tachyon matter is converted into a dark energy-like component (the tension of D-anti-D brane pairs), $\omega_T\rightarrow -1$. In D-brane scenario, it corresponds the re-production of D-anti-D-brane pairs by the effective vacuum fluctuations. Therefore, after the tachyon matter dominated contraction, a deflation driven by the tension of D-anti-D-brane pairs ensues. We call the phase transition from $\omega_T\rightarrow 0$ to $\omega_T\rightarrow -1$ as Reversal Tachyon Condensation. And we are discussing this transition in detail at next section.

\subsection{The Reversal of Tachyon Condensation Transition}
\label{sec-rtc}
Essentially, the reversal tachyon condensation transition is a reverse process of the coupled tachyon condensation. During the coupled tachyon condensation, tachyon rolls down from the top of its potential, and D-anti-D-brane pairs annihilate. Universe becomes tachyon matter dominated while EoS of tachyon field evolves from $\omega_T\rightarrow -1$ to $\omega_T\rightarrow 0$ as shown in~\refig{swTe.pdf}~. Accordingly, for reversal tachyon condensation transition, tachyon is pushed back to the top of its potential hill, $T\sim 0$ and $\rho_T\sim V_0$, by the blue-shifted $\phi$ during the contracting phase. In D-brane scenario, it indicates that D-anti-D-brane pairs are re-produced by the effective vacuum fluctuations. Universe is dominated by the tension of these brane pairs again. And the EoS of tachyon field evolves from $\omega_T\rightarrow 0$ to $\omega_T\rightarrow -1$ as shown in~\refig{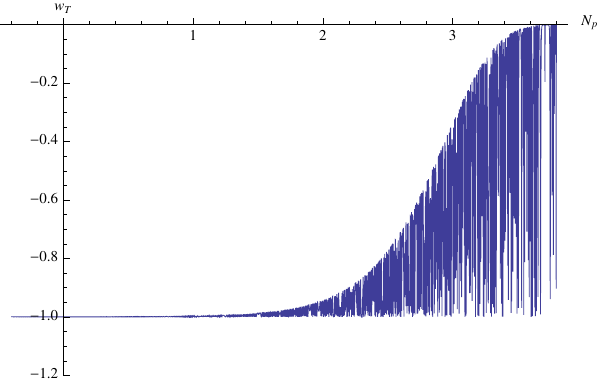}. Therefore, after reversal tachyon condensation, universe undergoes a deflation,
\begin{equation}
a=a_{ec}e^{-\sqrt{V_0}\tilde{M}_p^{-1}t}~, \label{eq-defl}
\end{equation}
driven by the tension of D-anti-D-brane pairs, $\omega_T\sim -1$, according to \refeq{Friedmann}~. And in \refeq{defl}~, we have neglected the sub-dominated energy density of $\phi$ and curvature term.

Now, utilizing the setup of numerical simulation presented in~\refsec{nsa}~, we are performing a numerical simulation for the evolution of the tachyon field, energy density and EoS of tachyon field in the coupled tachyon condensation and the reversal tachyon condensation respectively.

{\bf Coupled Tachyon Condensation:} For the coupled tachyon condensation, the evolution of tachyon field, EoS and energy density of tachyon field are plotted in the \refig{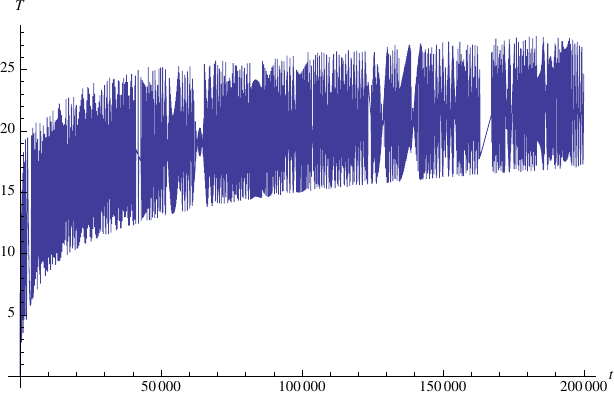}~, \refig{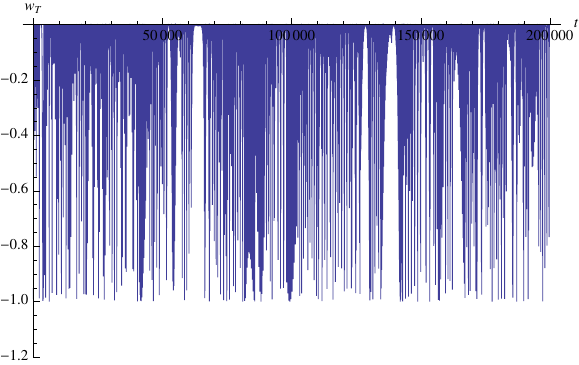} and \refig{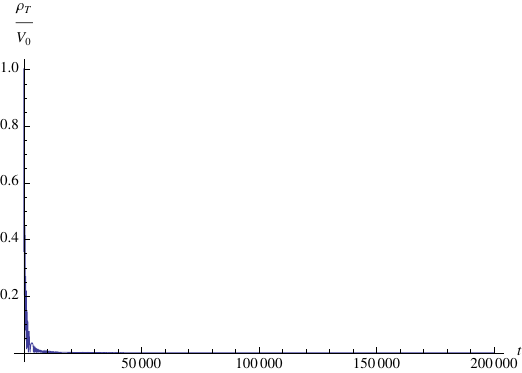}; and \refig{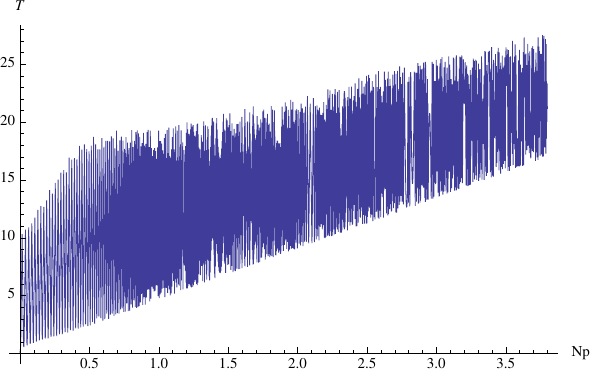}~, \refig{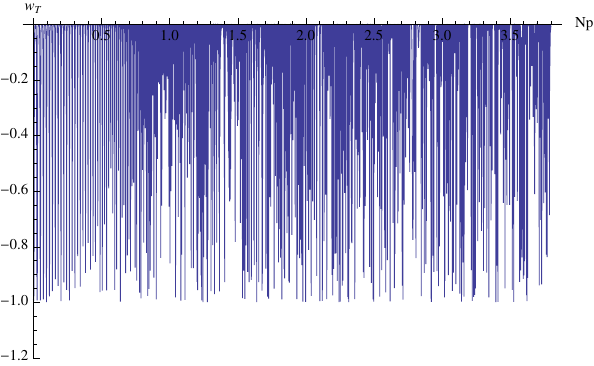} ,  and \refig{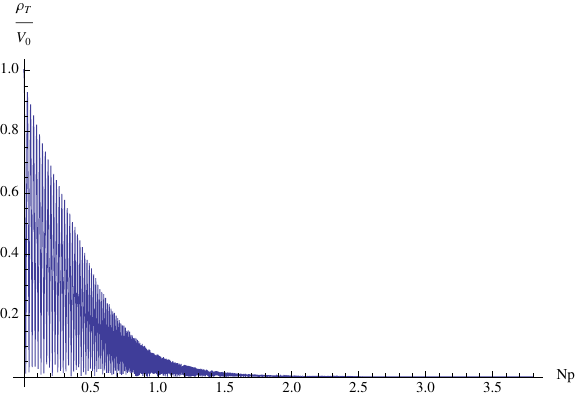}~ with respect to time and the e-folding number of scale factor respectively. For these numerical simulations, the initial conditions are taken as: $T[0]=0.1$, $\dot{T}[0]=0.1$, $L=\frac{2}{3}*10^3$, $k=0.1$, $t_i=0$, and $t_f=10^4$~.

During the coupled tachyon condensation, as shown in the~\refig{Te.pdf},  the expectation value of tachyon field increases with time, and the tachyon field oscillates around such expectation value very swiftly. In~\refig{Ten.pdf}, we find tha the expectation value of the tachyon field is linear to the e-folding number of scale factor, $T_v\propto N_p$, which is consistent to our analytical analysis in previous section. Moreover, both of ~\refig{wTe.pdf} and~\refig{wTen.pdf} show that, the equation of state of tachyon field evolves from $\omega_T\sim -1$ to $\omega_T\sim 0$ (even though it occasionally reaches $-1$, the equation of state of tachyon field generally stays around $0$) in the coupled tachyon condensation phase, see~\refig{swTe.pdf} for high resolution plotting. It indicates that the energy density of tension of D-anti-D-brane pairs releases into tachyon matter during the coupled tachyon condensation. And, in~\refig{rhoTe.pdf} and~\refig{rhoTen.pdf}, they show that the energy density of tachyon field decreases while the tachyon matter is diluted during this expanding phase. In summary, after the coupled tachyon condensation, universe evolves into the tachyon matter dominated epoch as discussed in \refsec{turnaround}~.

{\bf Reversal Tachyon Condensation:} For the coupled tachyon condensation, the evolution of tachyon field, EoS and energy density of tachyon field are plotted in the \refig{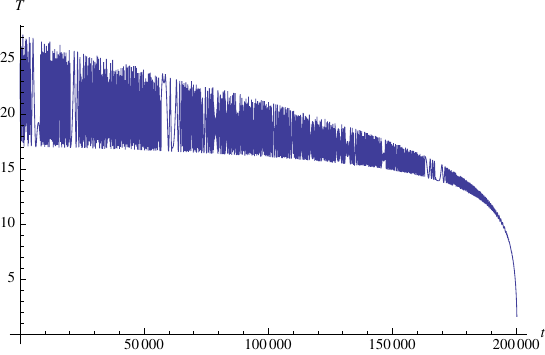}~, \refig{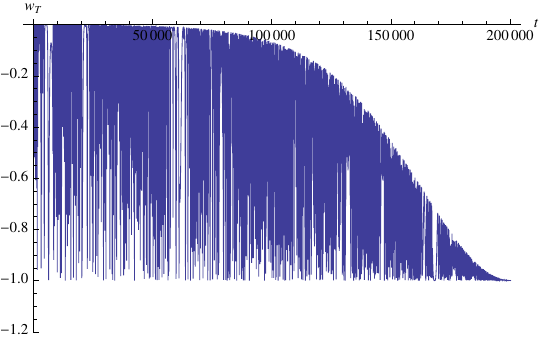} and \refig{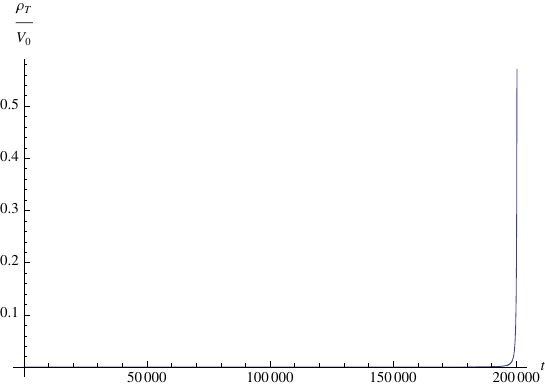}; and \refig{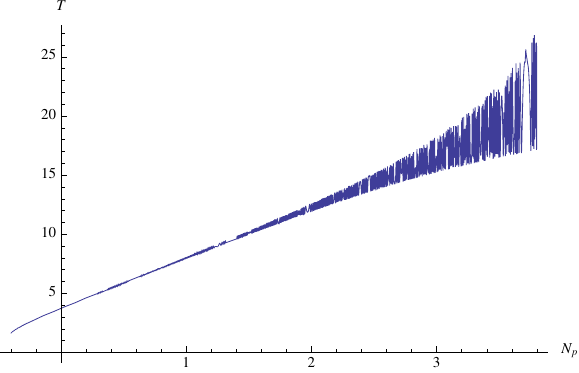}~, \refig{wTcn.pdf}  and \refig{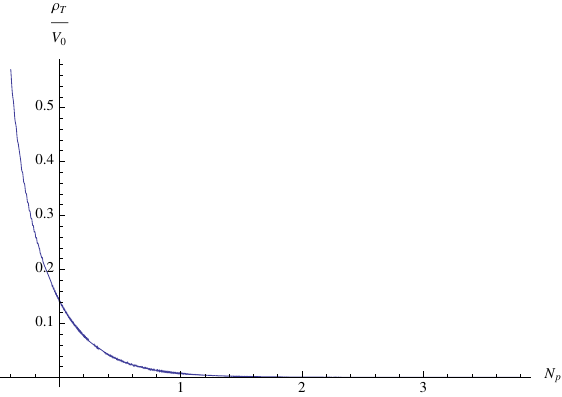}~,  respectively, with respect to time and the e-folding number of scale factor.  Accordingly, the initial conditions are taken as: $T[t_{ic}]=14.2303$, $\dot{T}[t_{ic}]=0.999561$, $L=\frac{2}{3}*10^3$, $k=0.1$, $t_{ic}=\frac{a_{max}}{a_1}L-(L+t_f)$ and $t_{fc}=\frac{a_{max}}{a_1}L-0.55L$ for the reversal tachyon condensation~\footnote{
To make the numerical simulation accessible in this tachyon matter dominated contraction, we take a cut-off, $0.55L$, in $t_{fc}$. It renders that each plotting in \refig{Tc.pdf}~, \refig{wTc.pdf} and \refig{rhoTc.pdf} moves to right side while the plots in \refig{Tcn.pdf}~, \refig{wTcn.pdf}  and \refig{rhoTcn.pdf} move to left side systematically and slightly. In principle, it would not affect the generic features of each plotting.
}. We notice that, since $N_p$ decreases in the contraction phase, each quantity, which is plotted with respect to e-folding number of scale factor, evolves from right side to left side in \refig{Tcn.pdf}~, \refig{wTcn.pdf} , and \refig{rhoTcn.pdf}~ respectively.

At the beginning of the reversal tachyon condensation, universe undergoes a tachyon matter dominated contraction. As shown in \refig{Tc.pdf} and \refig{Tcn.pdf}~, the expectation value of tachyon field, $T_v$, evolves toward $0$, and the amplitude of the oscillations of tachyon field around the effective vacuum also decreases to be $0$. Especially, in the~\refig{Tcn.pdf}, we find that the expectation value of the tachyon field  is linear to the e-folding number of scale factor, $T_v\propto N_p$. Again, it is consistent to our analytical analysis in previous section. In \refig{wTc.pdf} and \refig{wTcn.pdf}~, they show that, during the reversal tachyon condensation, the equation of state of tachyon field evolves from $\omega_T\sim 0$ to $\omega_T\sim -1$. It indicates that the energy density of tachyon matter is converted into a dark energy-like component (the tension of D-anti-D-brane pairs reproduced by the effective vacuum fluctuations) after the reversal tachyon condensation. In \refig{rhoTc.pdf} and \refig{rhoTcn.pdf}~, accordingly, the energy density of tachyon field increases to the order of $V_0$ when the tachyon field is pushed back at the end of this reversal tachyon condensation phase.

To sum up, after the reversal tachyon condensation, universe undergoes a deflation, \refeq{defl}~, driven by the tension of reproduced D-anti-D brane pairs until the curvature term being co-dominated. Then universe is bouncing from the deflation to the locked inflation of next cycle in competition of curvature term and tension of D-anti-D brane pairs. The details of such smooth bounce are discussed in next section.

\fig{Te.pdf}{The evolution of the tachyon field is plotted with respect to time in the coupled tachyon condensation. The expectation value of tachyon field increases, and the tachyon field oscillates around its expectation vacuum, $T_v$, swiftly.}{0.70}{h!t}

\fig{Ten.pdf}{The evolution of the tachyon field is plotted with respect to the e-folding number of scale factor, $N_p$,  in the coupled tachyon condensation. The expectation value of the tachyon field increases and is linear to the e-folding number, $T_v\propto N_p$. And tachyon field oscillates around such expectation vacuum swiftly.}{0.70}{h!t}

\fig{wTe.pdf}{The evolution of the equation of state of tachyon field is plotted with respect to time in the coupled tachyon condensation. Even though it occasionally reaches $-1$, the equation of state of tachyon field generally stays around $0$ in this phase.}{0.70}{h!t}

\fig{wTen.pdf}{The evolution of the equation of state of tachyon field is plotted with respect to the e-folding number of scale factor, $N_p$, in the coupled tachyon condensation. Even through it occasionally reaches $-1$, the equation of state of tachyon field generally stays around $0$.}{0.70}{h!t}

\fig{rhoTe.pdf}{The energy density of tachyon field is plotted with respect to time in the coupled tachyon condensation.  }{0.70}{h!t}

\fig{rhoTen.pdf}{The energy density of tachyon field is plotted with respect to the e-folding number of the scale factor, $N_p$, in the coupled tachyon condensation. }{0.70}{h!t}


\fig{Tc.pdf}{The evolution of tachyon field is plotted with respect to time in the reversal tachyon condensation. The expectation value of tachyon field, $T_v$, evolves toward $0$ while the amplitude of the oscillations of tachyon field around the effective vacuum also decreases to be $0$.}{0.70}{h!t}

\fig{Tcn.pdf}{The evolution of the tachyon field is plotted with respect to the e-folding number of scale factor, $N_p$, in the reversal tachyon condensation ({\bf Right $\rightarrow$ Left}). The expectation value of tachyon field, $T_v$, evolves toward $0$, and the amplitude of the oscillation of tachyon field around the effective vacuum also decrease to $0$. Moreover, the expectation value of the tachyon field is linear to the e-folding number, $T_v\propto N_p$, in this contracting phase.}{0.70}{h!t}

\fig{wTc.pdf}{The equation of state of the tachyon field is plotted with respect to time in the reversal tachyon condensation. During this contracting phase, the value of equation of state of tachyon field, $\omega_T$, evolves towards $-1$ from $0$.}{0.70}{h!t}

\fig{wTcn.pdf}{The equation of state of the tachyon field is plotted with respect to the e-folding number of scale factor, $N_p$, in the reversal tachyon condensation ({\bf Right $\rightarrow$ Left}). During this contracting phase, the value of equation of state of tachyon field, $\omega_T$, evolves towards $-1$ from $0$.}{0.70}{h!t}

\fig{rhoTc.pdf}{The energy density of the tachyon field is plotted with respect to time in the reversal tachyon condensation. The energy density of tachyon field increases to the order of $V_0$ when the tachyon field is pushed back at the end of this contracting phase.}{0.70}{h!t}

\fig{rhoTcn.pdf}{The energy density of the tachyon field is plotted with respect to the e-folding number of scale factor, $N_p$, in the reversal tachyon condensation ({\bf Right $\rightarrow$ Left}). The energy density of tachyon field increases to the order of $V_0$ when the tachyon field is pushed back at the end of this contracting phase.}{0.70}{h!t}

\subsection{Smooth Bounce and Cyclic Cosmic Evolution}

After the reversal tachyon condensation, universe is dominated by the dark energy-like component, {\it i.e.} the tension of D-anti-D brane pairs, $\rho_T=V_0$~. Taking account into the curvature term,
\begin{equation}
(\frac{\dot{a}}{a})^2=-\frac{1}{a^2}+\frac{V_0}{\tilde{M_p}^2}~,
\end{equation}
universe undergoes a smooth bounce,
\begin{equation}
a=a_\ast\cosh[\sqrt{\frac{V_0}{\tilde{M_p}^2}}(t-t_\ast)]~,\label{eq-abc}
\end{equation}
where $t_\ast=t_f+\sqrt{V_0}\tilde{M}_p^{-1}arc\cosh[a_fa_\ast^{-1}]$ and $a_\ast=\sqrt{V_0}\tilde{M}_p^{-1}$. Here, we use the subscripts, $_f$ and $\ast$, to denote the quantities at the end of the reversal tachyon condensation and at the bounce point respectively. At $t=t_\ast$~, universe reaches its non-zero minimum, $a=a_\ast=\sqrt{V_0}\tilde{M}_p^{-1}$, and is bouncing from the deflation phase to a new inflation phase. Clearly, such smooth bounce connects the deflation, \refeq{loin}~, and the locked inflation of new cycle, \refeq{defl}~. It indicates that, up to the zeroth order of cosmological evolution, universe driven by the coupled scalar-tachyon fields undergoes a cyclic evolution. In summary, with the smooth bounce, universe driven by the coupled scalar-tachyon fields is free of the Big-bang singularity.

One may be worried about that, during the tachyon matter dominated contraction or the deflation phase, the energy density of scalar field might become dominated by strongly blue-shifted, which renders the smooth bounce invalid.  Now we demonstrate that the energy density of tachyon field would dominate over that of $\phi$ during the whole process. According to the previous analytical analysis and numerical simulations, the e-folding number of locked inflation, $N_i$, is almost equal to that of locked deflation, $N_d$, $N_i\simeq -N_d$. And the e-folding number of the tachyon dominated expansion, $N_e$, is also almost equal to that of the contraction phase, $N_c$, $N_e\simeq -N_c$.  During the  locked inflation and the deflation,  $\rho_\phi=m^2\langle\phi^2\rangle\propto e^{-3N}$~.  And during the tachyon matter dominated expansion and the contraction, $\rho_\phi=(m^2+2\lambda T^2)\langle\phi^2\rangle\propto T^2e^{-3N}T^{-1}\propto f (3N+f^{-1})*e^{-3N}$, where $f$ is a constant determined by the relation $\sqrt{2} M_s (3N_p+f)=T_v$~.  Straightforwardly, we have
\begin{enumerate}
\item At the ending of locked inflation phase, $\rho_{\phi e}=e^{-3N_i}\rho_{\phi 0}$;

\item At the turnaround point, $\rho_{\phi t}=f*3N_e*e^{-3N_e+f^{-1}}\rho_{\phi e}$ ;

\item At the ending of tachyon matter dominated contraction , $\rho_{\phi c}=f*3(N_e+N_c+f^{-1})*e^{-3(N_e+N_c)}\rho_{\phi e}$~;

\item At the bouncing point, $\rho_{\phi \ast}=e^{-3N_d}\rho_{\phi c}=f*3(N_e+N_c+f^{-1})*e^{-3(N_e+N_c)}*e^{-3(N_i+N_d)}\rho_{\phi 0}$
\end{enumerate}
With the conditions,  $N_i\simeq -N_d$, $N_e\simeq -N_c$, we have $\rho_{\phi \ast}\simeq\rho_{\phi 0}\ll V_0$. We find that the energy density of scalar field is always sub-dominated comparing with tachyon field during the contracting phase and deflation phase.

\section{Conclusion}
\label{sec-concl}
In this paper, we present a string-inspired coupled scalar-tachyon fields model for a bouncing/cyclic universe without initial singularities in the closed FLRW background. A scalar-tachyon coupling term have been introduced into the action to make the coupled tachyon condensation come out naturally after the locked inflation driven by the tension of D-anti-D brane pairs. Further investigation reveals that such scalar-tachyon coupling term plays a crucial role in cosmic evolution of universe, and distinguishes the coupled scalar-tachyon model from other traditional single tachyon field cosmology and/or D-brane inflation cosmology.

We take an detailed analytical analysis and numerical simulation to study the entire cosmology evolution of this model. The cosmological evolution mainly consists of six distinctive phases: locked inflation, coupled tachyon condensation, tachyon matter dominated era(expansion, turnaround and contraction), reversal tachyon condensation, deflation and smooth bounce. Each phase has been studied analytically and numerically in this paper. No ghosts are ever generated at any point in the entire evolution of the universe. And the Null, Weak, and Dominant Energy Conditions are not violated at the bounce points. Moreover, the smooth bounce connects the deflation and the locked inflation of a new cycle. It indicates that, up to the zeroth order of background, universe enjoys a cyclic evolution. 

To make contact with the cosmological observations, we have studied the density perturbations of this model in~\cite{Li:2012vi} and~\cite{Li:2013bha}. And we find that, for this model, the power spectrum of curvature perturbation is nearly scale-invariant, $n_s-1 \simeq 0$, in consistent with recent observations~\cite{Komatsu:2010fb,Planck:2013kta}.

\section*{Acknowledgments}

We wish to thank  Yifu Cai,  Anke Knauf, Yi Wang,   Mingzhe Li,  and Konstantin Savvidy for many useful discussions and comments. Enlightening discussions with Feng Xu, Zhen Yuan and Yun Zhang at various stages of the work are also gratefully acknowledged.  We  would like to thank various KITPC programs where this work has been presented.  Last but not the least E.~C. enjoyed the wonderful hospitality and  discussions with  Ian McArthur at University of Western Australia, Larus Thoracius and Marcus Berg at Nordita, Stephen Hwang at  Linnaeus University. 

This research work has enjoyed supports, in parts,   from  
 the Jiangsu Ministry of Science and Technology under contract~BK20131264, the NSFC grant No.~10775067,  and by the Swedish Research Links programme of the Swedish Research Council (Vetenskapsradets generella villkor) under contract~348-2008-6049. 

We also acknowledge
985 Grants from the Ministry of Education, and the
Priority Academic Program Development for Jiangsu Higher Education Institutions (PAPD).

\appendix
\secs{The Setup of Numerical Simulation}
\label{sec-nsa}

After the locked inflation, the equation of motion of tachyon field can be much simplified for the numerical simulations. Here we set up the numerical equations for the simulations of these processes after the locked inflation. 

First of all, \refeq{EoM-phi} can be rewritten as
\begin{equation}
\ddot{\psi}+(m_{e\phi}^2-\frac{9p}{2\tilde{M_p}^2})\psi=0, \quad \phi(t) \equiv a^{-\frac{3}{2}}\psi(t)~, \label{eq-ddpsi}
\end{equation}  
where $p$ is a constant, $\tilde{M}_p\equiv\sqrt{\frac{3}{8\pi}}M_p$ and $m_{e\phi}^2=\lambda T^2+m^2\simeq \lambda T^2 \gg\frac{9p}{2M_p^2}$~.~With the ansatz, $T=h*t$, the solution of \refeq{ddpsi} is
\begin{equation}
\psi\propto({h*t})^{-\frac{1}{2}}\times\cos[\sqrt{\lambda} ht^2],\quad h*t\gg 0~.
\end{equation} 
Integrating the fast oscillating factor $\cos[\sqrt{\lambda} h t^2]$ out,  $\langle\psi^2\rangle\propto({h*t})^{-1}$, where $\langle X \rangle$ denotes the averaged value of $X$. Then we obtain
\begin{equation}
{\langle \phi^2\rangle} T\propto a^{-3}~.\label{eq-phist}
\end{equation} 
Substituting \refeq{phist} into \refeq{EoM-T}~, the equation of motion of tachyon field is simplified,
\begin{equation}
\ddot{T}+(1-\dot{T}^2)[-\frac{1}{\sqrt{2}}\tanh(\frac{T}{\sqrt{2}})+k
e^{-3N_p}\sqrt{1-\dot{T}^2}\cosh(\frac{T}{\sqrt{2}})]=0~, \label{eq-sddT}
\end{equation}
where  $N_p$ is the e-folding number of scale factor after the locked inflation, $N_p\equiv \ln(\frac{a}{a_{1}})$, $\gamma$ is a numerical parameter determined by the initial conditions, and $k\equiv \frac{1}{2}\frac{\phi^2_1}{\phi_c^2}\gamma$~.  With $H\sim \frac{1}{\sqrt{M_p}}$, the term, $3(1-\dot{T}^2)H\dot{T}$, is much smaller than others, $\frac{1}{\sqrt{2}}\tanh(\frac{T}{\sqrt{2}})$ and $ke^{-3N_p}\sqrt{1-\dot{T}^2}\cosh(\frac{T}{\sqrt{2}})$,  and has been neglected above. In \refeq{sddT}~, we use subscript $_1$ to denote the quantities at end of locked inflation, and adopt the unit convention, $M_s=10^{-3}\tilde{M}_p=1$~.

After the locked inflation, universe evolves into a tachyon matter dominated era with three phases, expansion, turnaround and contraction. Here we list the equations of motion of each phase according to \refeq{sddT}, where $L\equiv \frac{2}{3}\frac{\tilde{Mp}}{\sqrt{V_0}}$,
\begin{enumerate}
\item In tachyon matter dominated expansion,  $a=a_1[\frac{3}{2}\frac{\sqrt{V_0}}{\tilde{M_p}}(t+\frac{2}{3}\frac{\tilde{Mp}}{\sqrt{V_0}})]^\frac{2}{3}$, 
\begin{equation}
\ddot{T}+(1-\dot{T}^2)[-\frac{1}{\sqrt{2}}\tanh(\frac{T}{\sqrt{2}})+k\frac{L^2}{(t+L)^2}
\sqrt{1-\dot{T}^2}\cosh(\frac{T}{\sqrt{2}})]=0~, \label{eq-sddTe}
\end{equation}
\item Around the turnaround point, 
\begin{equation}
\ddot{T}+(1-\dot{T}^2)[-\frac{1}{\sqrt{2}}\tanh(\frac{T}{\sqrt{2}})+k\frac{a_1^3}{a_{max}^3}
\sqrt{1-\dot{T}^2}\cosh(\frac{T}{\sqrt{2}})]=0~, \label{eq-sddTt}
\end{equation}
\item In tachyon matter dominated contraction, $a=a_1[\frac{2}{3}\frac{\sqrt{V_0}}{\tilde{M_p}}(-t+\frac{2}{3}\frac{\tilde{M_p}}{\sqrt{V_0}}\frac{a_{max}^3}{a_1^3})]^\frac{2}{3}$,
\begin{equation}
\ddot{T}+(1-\dot{T}^2)[-\frac{1}{\sqrt{2}}\tanh(\frac{T}{\sqrt{2}})+k\frac{L^2}{(\frac{a_{max}^3}{a_1^3}L-t)^2}
\sqrt{1-\dot{T}^2}\cosh(\frac{T}{\sqrt{2}})]=0~.\label{eq-sddTc}
\end{equation}
\end{enumerate}

\section{Parameters Analysis}
\label{sec-Analysis}

\ssecs{Constraints from Each Phase of Universe Evolution}
\label{ssec-Constraint}
Let's first list the constraints of the model which we have to take into account when computing the e-folding.   We will use subscript 0 to denote initial value at the bounce point. Each constraint in this section is given  followed with explanations.

\cnstr{Vacuum Energy Domination}{
\frac{V_0}{2\lambda\ta{T_0^2}\ta{\phi_0^2}}>1&b,c,e,f\l
\frac{V_0}{m^2\ta{\phi_0^2}}>2&a,d\label{eq-con1}}

The first constraint comes from our expectation of initial tachyon (or its vacuum energy) domination. Not only  it is the requirement from locked inflation,  it is also because the total equation of state should be smaller than $-1/3$, that of curvature.   Otherwise, the universe would contract instead of expand. Depending on which term is larger in the mass square of $\phi$, two cases are considered separately to get~\refeq{con1}.

\cnstr{Initial Locking}{
\ta{\phi_0^2}>\ta{\phi_c^2}&a,b,c\l
m^2\ta{\phi_0^2}>\frac{V_0}{2\ms^2}\ta{\phi_c^2}&d\l
2\lambda\ta{T_0^2}\ta{\phi_0^2}>\frac{V_0}{2\ms^2}\ta{\phi_c^2}&e,f.}

We would like to have a period  of locked inflation, and we would like  it to make a major contribution to  e-folding.  This  constraint is to ensure initial tachyon locking, by making $\ta{\phi_0^2}$ larger than the $\ta{\phi^2}$ at the end of locked inflation. We categorize cases $a$ to $f$ according to different ending conditions of locked inflation, the definitions and therefore the values of $\ta{\phi_1^2}$ are also different in  different cases.

\cnstr{Fastroll Scalar Field}{%
m^2>\frac{6\pi}{\mpl^2}(V_0+m^2\ta{\phi_0^2})  &   a,d \l 
2\lambda\ta{T_0^2}>  \frac{6\pi}{\mpl^2}  
(V_0+2\lambda\ta{T_0^2} \ta{\phi_0^2})      &   b,c,e,f\l
\mpl^2\ta{T_0^2}>12\pi\ms^2\ta{\phi_0^2}    &  b\l
m^2\mpl^4\ta{T_0^2}>144\pi^2\lambda V_0 \ms^2 \ta{\phi_0^2}   &   c\l
\mpl^4\ta{T_0^2}  > 288\pi^2 \lambda \ms^4 \ta{\phi_0^2}   &    e,f.
}

To get an oscillating $\phi$ during locked inflation, we need $\phi$ to violate the slow-roll condition. The effective mass of $\phi$ thus yields $m_\phi^2>9H^2/4$. Neglecting  the smaller contribution to  
$\phi$'s mass, the derivation of above constraint is straightforward. The first two relations are for initial fast-roll property and the rest are for fast-roll near the end of locked inflation. 

\cnstr{Proper Mass of Scalar Field}{
m^2>2\lambda\ta{T_0^2}&a,d\l
m^2<\frac{V_0\ta{T_0^2}}{2\ms^2\ta{\phi_0^2}}&b\l
2\lambda\ta{T_0^2}>m^2>\frac{V_0\ta{T_0^2}}{2\ms^2\ta{\phi_0^2}}&c\l
m^2<\frac{V_0}{2\ms^2}\sqrt\frac{\ta{T_0^2}}{\ta{\phi_0^2}}&e\l
2\lambda\ta{T_0^2}>m^2>\frac{V_0}{2\ms^2}\sqrt\frac{\ta{T_0^2}}{\ta{\phi_0^2}}&f.}

These and the next constraints come from the defining characteristics of each case. 
The above  constraints ensure the correct relationship between $\lambda\ta{T_0^2}$, $\lambda\ta{T_1^2}$ and $m^2$. The next ones apply to $m_\phi^2$ and $V_0/2\ms^2$ at Point 1. 

\cnstr{Ending Condition of Locked Inflation}{
2m^2\ms^2>V_0&a,c\l
\ta{T_0^2}>\ta{\phi_0^2}&b\l
2m^2\ms^2<V_0&d,f\l
\ta{T_0^2}<\ta{\phi_0^2}&e.}

We can see from this set of  constraints $\sqrt{V_0/2\ms^2}$ is a critical value of $m$ which divides the moduli space into normal ending and mass ending for $m$ that is not too small ($m^2>2\lambda\ta{T_1^2}$). This is reasonable because as long as $m$ is not too small, it would already be $m$ dominating in the mass of $\phi$, so the constraint only applies to $m$. Moreover, after checking with other constraints we  find that  the case $e$ may only exist under $m^2<V_0/2\ms^2$. So in the analysis of moduli space in the next section, we will discuss the large $m$ and small $m$ separately.

\cnstr{Relaxed Tachyon}{
\lambda m^2\ms^4\mpl^2\ta{\phi_0^2}^2>3\pi V_0^3&a,d\l
2\lambda^2\ms^4\mpl^2\ta{\phi_0^2}^2\ta{T_0^2}>3\pi V_0^3&b,c,e,f.}

This is the last constraint, coming  from the number of e-folds. We expect the universe to be still expanding when $T=T_V$ at Point 3, so there would be reheating and baryon genesis for late time universe. According to the evolution of the ratio of positive energy density to curvature energy, we write the constraint as $2(N_L+N_R)>N$, based on that during inflation energy density stays constant, and afterwards the universe becomes matter-like.

\ssecs{Moduli Space and E-folds}
\fig{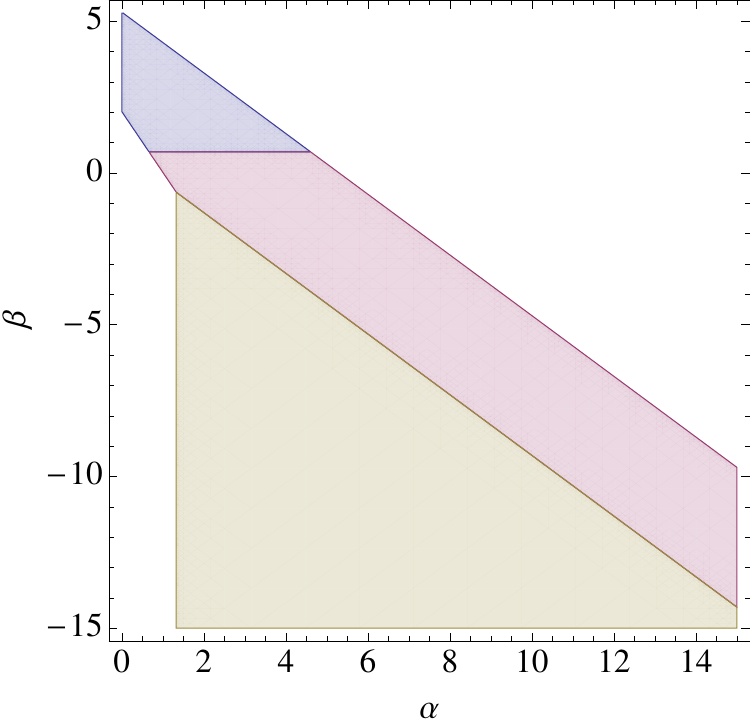}{The moduli space diagram for $m^2>V_0/2\ms^2$, when $m^2=10^{-3}\mpl^2$, $\ta{\phi_0^2}=10^{-8}\mpl^2$. Brown, blue and red correspond to cases $a,b,c$ respectively.}{0.6}{}
\fig{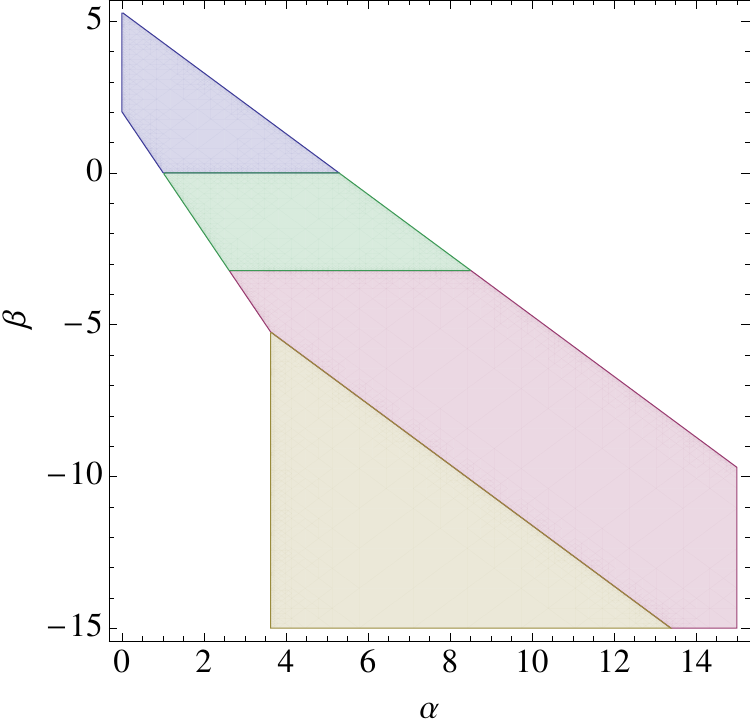}{The moduli space diagram for $m^2<V_0/2\ms^2$, when $m^2=10^{-4}\mpl^2$, $\ta{\phi_0^2}=10^{-8}\mpl^2$. Blue, brown, green and red correspond to $a,d,e,f$.}{0.6}{}

The string mass scale, $\ms$, is usually chosen to be $10^{-3}\mpl$, the scale at which  we are also adopting here. $V_0$ is the scale of $\ms^3\mpl$ from definition of tachyon in string theory. Therefore we have four free parameters of the system, $\lambda$, $m$, $\ta{T_0^2}$ and $\ta{\phi_0^2}$. Once we give them definite values in the moduli space, the evolution of the universe is fully determined. For demonstration, we choose
\eqa{
\alpha&\equiv& \ln\frac{4\lambda\ms^2\ta{\phi_0^2}}{V_0},\\
\beta&\equiv&\ln\frac{\ta{T_0^2}}{\ta{\phi_0^2}}}
as the axes of moduli space, and use $\ta{\phi_0^2}$ and $m^2$ as the free parameters. Any combination of values of $\ta{\phi_0^2}$ and $m^2$ provides a different moduli space of $\alpha$ and $\beta$.

Such a choice of axes and free parameters provides convenience in several aspects. $\alpha$ is actually $\ln\ta{\phi_0^2}/\ta{\phi_c^2}$ so it partly indicates the e-folds of locked inflation. Definition of $\beta$ simplifies the  constraints, especially~\refcn{Proper Mass of Scalar Field}. $m$ acts as a free parameter and provides distinctive moduli spaces under different values, as discussed in~\refcn{Ending Condition of Locked Inflation}.

Given the constraints in~\refssec{Constraint}, we can plot the moduli spaces. \refig{Moduli_Space1.pdf} is the case when $m^2>V_0/2\ms^2$, and~\refig{Moduli_Space2.pdf} is for $m^2<V_0/2\ms^2$. From these two figures, we can tell that if we want to increase the e-folds of locked inflation (characterized by $\alpha$), we need to decrease $\beta$ (i.e. $\ta{T_0^2}$, since $\ta{\phi_0^2}$ remains constant as a free parameter) to preserve tachyon domination(\refcn{Vacuum Energy Domination}). Other constraints can be found in the figures in the same way.

\fig{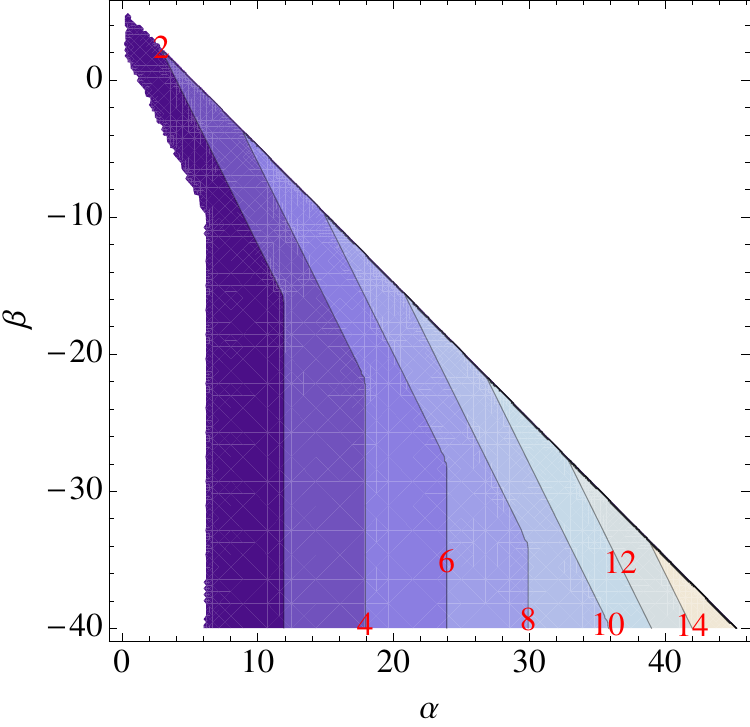}{The number of remaining e-folds is shown by the contours, with the specific numbers in red. We have taken the parameter values $m^2=10^{-5}\mpl^2$, $\ta{\phi_0^2}=10^{-8}\mpl^2$~.}{0.6}{}
\fig{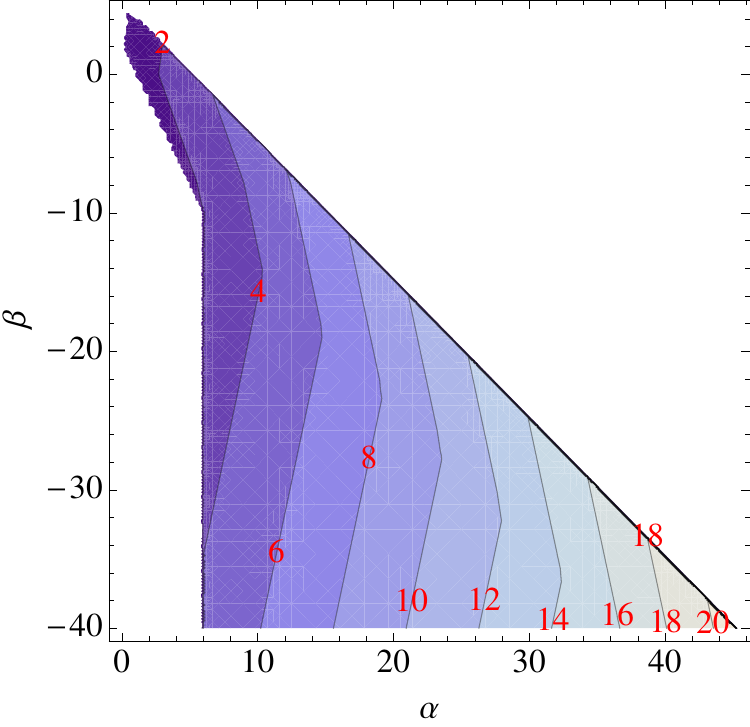}{The total number of e-folds of inflation is shown by the contours, with the specific numbers in red. We have taken the parameter values $m^2=10^{-5}\mpl^2$, $\ta{\phi_0^2}=10^{-8}\mpl^2$~.}{0.6}{}
We can then further demonstrate the number of e-folds within the moduli space;~\refig{E-folds_Left.pdf} gives the figure of remaining e-folds after $T_3=T_{V3}$ before turnaround, which is defined as $2(N_L+N_R)-N$. And~\refig{E-folds_Inflation.pdf} shows the total number of e-folds of inflation $N_L+N_R$. It can be seen that in most cases, locked inflation indeed contributes much more to total e-folds than rolling inflation. On the other hand, decreasing $\beta$ does generate additional e-folds that can contribute a small portion on total e-folds in~\refig{E-folds_Inflation.pdf}. The vertical contours in~\refig{E-folds_Left.pdf} reveals that the e-folds required by the catch-up process  of $T$ with its \vev{} $T_V$ cancels that from  rolling inflation.

If we want the e-folds of locked inflation to be  $N_L=50$, there would be $\alpha\approx 150$ and $\beta<\sim-150$. The large $\alpha$ would imply a strong coupling between $\phi$ and $T$, which means tachyon will not generate until branes become very close. Meanwhile, $\beta$ demands an initial $\ta{T^2}$ frozen at zero. This is reasonable because the initial large distance between the branes prevents tachyon production, and the strong coupling further ensures that.  Therefore, our model needs a strong coupling between tachyon and the distance between branes. Otherwise it is difficult to get an e-folding larger than 50.

\ssecs{Discussion of Parametric Resonance}
\label{sec-dpr}
In the  above analysis,  we have neglected,  for simplicity,  the effect of parametric resonance.  Here we briefly discuss  its effect in the locked inflation scenario.

Parametric resonance during locked inflation transfers energy from $\phi$ field to tachyon, pumping up $\ta{T^2}$. In the language of particle physics, tachyon particles are generated because of the interaction and the oscillation of $\phi$.  This process is  usually called ``preheating'' in many papers (see \cite{Allahverdi:2010xz} for a brief review). When  back-reactions of $T$ on $\phi$ are negligible, such preheating produces the number density of $T$ that is exponentially increasing for all modes below a certain momentum. Such exponential increase of $T$ would certainly ruin our model  if no precaution is taken.

There are usually three ways to deal with unwanted resonance, one of which  works for our model. Let us first see why the other two do not. First if the effective mass of $T$ from $\phi$ is much smaller than $T$'s bare mass, the effect of parametric resonance is negligible. It is certainly inapplicable to our model because the number of e-folding of locked inflation requires $\phi$ contribute an effective mass that is much larger than $T$'s bare mass, $\sqrt{V_0/2\ms^2}$. The second point  is that if $\phi$ is fast rolling but not too fast, i.e. in the range of $10H>\sim m_\phi>3H/2$, the Hubble damping due to expansion is even larger and cancels parametric resonance effect. So this mechanism  works well for our model during expansion but it must fail at the contraction phase, because contraction acts as a boosting effect. Altogether with parametric resonance, $T$'s amplitude grows even faster during contraction.

The method we are using is to allow the back-reaction of $T$ so that parametric resonance between $\phi$ and $T$ is in equilibrium. Therefore the energy transfer to and fro each other by  parametric resonance should be exactly equal, i.e. the  growing rate of $\phi$ is proportional to the effective mass of $T$ and vice versa. With these relations, we have 
 $m_\phi\ta{\phi_e^2}=m_T\ta{T_e^2}$, where subscript $e$ means equilibrium and the effective masses are taken to be  $m_\phi=m$, $m_T=\sqrt{\lambda\ta{\phi_e^2}}$. Since we require many e-folds during locked inflation,  $m_T\gg m_\phi$  and $\phi$ is  initially large. 
 We thus get $\ta{T_e^2}/\ta{\phi_e^2}\ll 1$ initially, which in turn implies that  the parametric resonance effect is negligible at the beginning. As $\ta{\phi^2}$ decreases due to inflation, this ratio will grow and become significant. It may become quite large near the end of locked inflation, and thereafter the stage of rolling inflation would be shortened or even bypassed.  Consequently, the kinetic-potential energy ratio at point 3 may decrease significantly. In such cases, the reheating efficiency from tachyons kinetic energy is lowered and may become insufficient.  This calls for other mechanisms for sufficient reheating, such as by taking into consideration $\phi$'s decay.

\bibliography{LWCbibref}

\end{document}